\documentclass[aps,prd,groupedaddress,nofootinbib,twocolumn]{revtex4-2}
\usepackage{graphicx,color,amsmath,mathtools,amssymb}
\usepackage{tcolorbox}
\usepackage{amsthm}
\usepackage{graphicx}
\usepackage{color,soul}
\usepackage{soul}
\usepackage{bm}
\usepackage{subfigure}
\usepackage{hyperref} 
\hypersetup{colorlinks=true,citecolor=blue,linkcolor=magenta,filecolor=magenta,urlcolor=blue,}

\begin{document}
	\title{Double-charm tetraquark under the complex scaling method}
	
	\author{Jian-Bo Cheng}%
	\email{jbcheng@pku.edu.cn} 
	\author{Zi-Yang Lin}
	\email{lzy\_15@pku.edu.cn}
	
	\author{Shi-Lin Zhu}
	\email{zhusl@pku.edu.cn}
	\affiliation{
		School of Physics and Center of High Energy Physics, Peking University 100871, China
	}%
	
	\date{\today}
	
	\begin{abstract}
		The LHCb Collaboration discovered a double-charm tetraquark $T_{cc}^{+}$ with a very small width. We investigate the $T_{cc}^{+}$ as a $DD^{*}$ molecule with $J^{P}=1^{+}$ in the framework of the one-boson-exchange potential model. The isospin breaking effect and $S-D$ wave coupling are taken into account carefully. We adopt the complex scaling method (CSM) to study the $DD^{*}$ system and obtain a quasibound state corresponding to the $T_{cc}^{+}$. Its binding energy relative to the $D^{0}D^{*+}$ and width are $-354$ keV and $61$ keV respectively. The isospin breaking effect is found to be enormous, and the $S-$wave $D^{0}D^{*+}$ and $D^{+}D^{*0}$ components give dominant contributions with the probabilities of $72.1\%$ and $27.1\%$ respectively. In addition, we do not find any resonances in the $DD^{*}$ system. As a by-product, we study the $X(3872)$ as a $(D\bar{D}^*-D^*\bar{D})/\sqrt{2}$ molecule with $J^{PC}=1^{++}$. We also find a quasibound state corresponding to the $X(3872)$. Its binding energy relative to the $D^{0}\bar{D}^{*0}$ threshold and width are $-111$ keV and $26$ keV respectively. The $S-$wave $(D^{0}\bar{D}^{*0}-D^{*0}\bar{D}^{0})/\sqrt{2}$ component dominates this state with the probability of $92.7\%$. 
	\end{abstract}
	\maketitle
	\section{Introduction}\label{sec:Introduction}
	Recently, the LHCb Collaboration reported the observation of a double-charm tetraquark $T_{cc}^+$ in the $D^0D^0\pi^+$ mass distribution \cite{lhcbcollaborationObservationExoticNarrow2021a}. The mass relative to the $D^{*+}D^0$ threshold and the width extracted from the Breit-Wigner fit are
	\begin{eqnarray}
		&&\delta m_{\text{BW}}=-273\pm61 \ \text{keV/c}^2\ \text{and}\nonumber\\
		&&\Gamma_{\text{BW}}=410\pm165 \ \text{keV}.\label{lhcb 1}
	\end{eqnarray}
	Obviously, the mass is just below the $D^{*+}D^0$ threshold, and the width is quite narrow, which indicates a $DD^{*}$ molecule structure. Subsequently, the LHCb Collaboration also analyzed the pole mass relative to the $D^0D^{*+}$ threshold and width in Ref. \cite{lhcbcollaborationStudyDoublyCharmed2021a}, and gave
	\begin{eqnarray}
		&&\delta m_{\text{pole}}=-360\pm40_{-0}^{+4} \ \text{keV/c}^2 \ \text{and}\nonumber \\
		&& \Gamma_{\text{pole}}=48\pm2_{-14}^{+0}\ \text{keV}.\label{lhcb 2}
	\end{eqnarray}
	
	The theoretical investigations on the double-charm tetraquark have been going on for 40 years \cite{aderNarrowHeavyMultiquark1982,ballotFourQuarkStates1983,zouzouFourquarkBoundStates1986a,hellerExistenceStableDimesons1987,carlsonStabilityDimesons1988,silvestre-bracSystematicsSystemsChromomagnetic1992,manoharExoticQQqqStates1993,brinkTetraquarksHeavyFlavors1998,cookLatticeStudyInteraction2002a,gelmanDoesNarrowTetraquark2003d,jancTccDDMolecular2004,vijandeDynamicalStudyMesons2006,ebertMassesTetraquarksTwo2007,vijandeExoticMesonmesonMolecules2009,yangWaveStateConstituent2009,diasRelationTccBb2011,ohkodaExoticMesonsDouble2012b,duExoticStates2013,liCoupledchannelAnalysisPossible2013,ikedaCharmedTetraquarksTcc2014,albuquerqueQCDSumRules2019a}.
	In 2017, the LHCb Collaboration observed the first double-charm hadron $\Xi_{cc}^{++}$ \cite{aaijObservationDoublyCharmed2017a}, and this discovery encourages the research on other double-heavy hadrons, especially the double-charm tetraquark \cite{luoExoticTetraquarkStates2017,mehenImplicationsHeavyQuarkdiquark2017a,fontouraProductionExoticTetraquarks2019a,xuPotentialsChiralEffective2019,francisEvidenceCharmbottomTetraquarks2019a,agaevStrongDecaysDoublecharmed2019a,tanSystematicsQQBarq2020,yangDoubleheavyTetraquarks2020,chengDoubleheavyTetraquarkStates2021}. Until very recently, the discovery of the $T_{cc}^+$ have ignited a new round of passion for the investigation of the double-charm tetraquark state \cite{chenPredictingAnotherDoubly2021,feijooMassDistributionProduction2021,flemingDecaysDifferentialSpectra2021,mengProbingLongrangeStructure2021b,mengRevisitIsospinViolating2021a,albaladejoTccCoupledChannel2022,chenDoublyheavyTetraquarkStates2022,daiPredictionNewStates2022,dengItsPartners2022,duCoupledchannelApproachIncluding2022,kePossibleMolecularStates2022,lingCanWeUnderstand2022,liuHolographicTetraquarksNewly2022,yanSubleadingContributionsDecay2022,chenSystematicsHeavyFlavor2021,dongSurveyHeavyHeavy2021,chenHeavyFlavorMolecular2022,dengDecodingDoubleHeavy2022}. Besides, one can get more information from the extensive reviews in recent years \cite{brambillaStatesExperimentalTheoretical2020,chenHiddencharmPentaquarkTetraquark2016,guoHadronicMolecules2018,liuPentaquarkTetraquarkStates2019,mengChiralPerturbationTheory2022,chenUpdatedReviewNew2022}.
	
	In the aforementioned theoretical studies, the molecule and compact tetraquark pictures obtain most of the attention. In the compact tetraquark picture \cite{zouzouFourquarkBoundStates1986a,silvestre-bracSystematicsSystemsChromomagnetic1992,brinkTetraquarksHeavyFlavors1998,gelmanDoesNarrowTetraquark2003d,vijandeDynamicalStudyMesons2006,vijandeExoticMesonmesonMolecules2009,duExoticStates2013,luoExoticTetraquarkStates2017,mehenImplicationsHeavyQuarkdiquark2017a,agaevStrongDecaysDoublecharmed2019a,tanSystematicsQQBarq2020,yangDoubleheavyTetraquarks2020,chengDoubleheavyTetraquarkStates2021,chenDoublyheavyTetraquarkStates2022,dengDecodingDoubleHeavy2022}, the double-heavy systems have a heavy diquark-antiquark symmetry, which can contribute a deep attractive force. Therefore, this scheme may generate a bound state relative to the corresponding dimeson threshold. On the other hand, when the molecule picture is taken into account, the one-boson-exchange (OBE) potential may also provide an attractive potential to form a bound state. For instance, the work in \cite{liCoupledchannelAnalysisPossible2013} gives a prediction on the molecule $T_{cc}^{+}$ with $I(J^P)=0(1^+)$ quantum number, and the binding energy is $-0.47$ MeV, which is very close to the experimental value. Besides, the recent work in \cite{chenPredictingAnotherDoubly2021}  considered the isospin breaking effect and predicted a new resonance $T_{cc}^{'+}$.
	
	In this work, we investigate the $T_{cc}^+$ and $X(3872)$ in the molecule picture. Indeed, these two exotic hadrons have some similar features. Besides the small binding energy and narrow width, one can get a quite similar OBE potential. However, they also have some evident differences. For instance, the $X(3872)$ has some hidden-charm decay patterns, such as the $J/\psi\rho$ and $\eta_c\omega$. By comparison, the $T_{cc}^+$ has only the open-charm decay channels $DD\pi$ and $DD\gamma$, which provides us convenience when investigating the dimeson molecule. 
	
	In fact, compared with the general system like the $BB^{*}$ or $BD^{*}$, the $DD^{*}$ has a unique behavior due to the decay process $D^{*}\to D\pi$. In the one-pion-exchange (OPE) effective potential, the $0$-th component of the transferred momentum can be slightly larger than the pion mass. This difference will induce an effect---the pole location of the OPE potential will move from the imaginary axis (general) to the real axis ($DD^{*}$) in the complex transferred momentum plane. In the Lippmann-Schwinger equation (LSE) or the momentum space Schr\"odinger equation, one will get an imaginary part and reach a pole when integrating the OPE potential along the real momentum axis. In other words, the calculation is divergent. Some works \cite{liIsospinBreakingCoupledchannel2012,chenPredictingAnotherDoubly2021} take the Cauchy principal (PV) value for the OPE potential to deal with this problem. However, we tend to get over it in a different way.
	
	In our framework, we use the complex scaling method (CSM) \cite{aguilarClassAnalyticPerturbations1971,balslevSpectralPropertiesManybody1971a} to study the $DD^{*}$ molecule system. We will retain the imaginary contribution from the OPE potential. After a complex scaling operation, the pole of OPE potential would be rotated from the real axis to the upper half of the momentum plane. The integral along the real momentum axis would bypass the pole and avoid divergence. In this way, we could solve the problem and get the bound states and resonances directly. As a byproduct, we also discuss the $X(3872)$ as a molecule with the $J^{PC}=1^{++}$, and consider the isospin breaking effect. 
	
	This paper is organized as follows. In Sec. \ref{sec:framework}, we will introduce our framework explicitly. In Sec. \ref{sec: potentials}, we present the effective potentials. In Sec. \ref{sec: results}, we solve the complex scaled Schr\"odinger equation and give the results of the $T_{cc}^{+}$ and $X(3872)$ by adopting the OBE potential. The last section \ref{sec:summary} is a brief summary.
	\section{Framework}\label{sec:framework}
	
	In this work, we assume that the $T_{cc}^+$ is a molecule with the quantum number $J^P=1^+$. The mass of the $T_{cc}^+$ is very close to the threshold of the $D^{0}D^{*+}$/$D^{+}D^{*0}$, and the interactions between these two channels are similar. Thus the isospin breaking effect should be considered. The masses of the $D^{(*)}$ mesons and exchanged light mesons are shown in Table \ref{tab: mass meson}. To deal with the isospin breaking effect, we take into account the channels $D^{0}D^{*+}(^3S_1,^3D_1)$ and $D^{+}D^{*0}(^3S_1,^3D_1)$, see Table \ref{tab: channel}. We do not consider the channels $D^*D^*(^3S_1,^3D_1,^5D_1)$. Their thresholds are evidently higher than the mass of the $T_{cc}^{+}$, indicating very small contributions. 
	
	\begin{table}[htbp]
	\begin{tabular}{cccc}\hline\hline
		Mesons&Mass(MeV)&Mesons&Mass(MeV)\\
		\hline
		$D^{0}$&1864.84&$\pi^{\pm}$&139.57\\
		$D^{+}$&1869.66&$\pi^{0}$&134.98\\
		$D^{*0}$&2006.85&$\eta$&547.86\\
		$D^{*+}$&2010.26&$\rho$&775.26\\
		$\sigma$&600&$\omega$&782.66\\ 
		\hline\hline
	\end{tabular}
	\caption{The masses of the charmed mesons and exchanged light mesons in the OBE potential.}\label{tab: mass meson}
\end{table}

	\begin{widetext}
		\begin{center} 
			\begin{table}[htbp]
				\renewcommand{\arraystretch}{1.8}{
					\setlength\tabcolsep{8pt}{
						\begin{tabular}{cccccc}\hline\hline
							\centering
							System&$J^{PC}$&$1$&$2$&$3$&$4$\\\hline
							$T_{cc}^{+}$&$1^+$&$D^{0}D^{*+}(^3S_1)$&$D^{0}D^{*+}(^3D_1)$&$D^{+}D^{*0}(^3S_1)$&$D^{+}D^{*0}(^3D_1)$\\ 
							$X(3872)$&$1^{++}$&$[D^{0}\bar{D}^{*0}](^3S_1)$&$[D^{0}\bar{D}^{*0}](^3D_1)$&$[D^{+}D^{*-}](^3S_1)$&$[D^{+}D^{*-}](^3D_1)$\\ 
							\hline\hline
				\end{tabular}}}
				\caption{The channels of the double-charm system $DD^*$ and hidden-charm system $[D\bar{D}^{*}]$ under the isospin breaking effect. We adopt the following shorthand notations for simplicity, $[D^{0}\bar{D}^{*0}]=\frac{1}{\sqrt{2}}(D^{0}\bar{D}^{*0}-D^{*0}\bar{D}^{0})$ and $[D^{+}\bar{D}^{*-}]=\frac{1}{\sqrt{2}}(D^{+}D^{*-}-D^{*+}D^{-})$.}\label{tab: channel}
			\end{table} 
		\end{center}
	\end{widetext}

	We first consider the OPE potential. As mentioned above, the $DD^*$ system is unique in contrast to the $\bar{B}\bar{B}^*$ or $\bar{B}D^*/D\bar{B}^*$ system---the $D^*$ can decay to $D\pi$, which provides an imaginary part in the OPE potential. To study the influence of the imaginary part, we could adopt one of the following equations: \\
	1) The coordinate space Schr\"odinger equation 
	\begin{eqnarray}
		\left[-\frac{1}{2m}\frac{d^2}{dr^2}+\frac{l(l+1)}{2mr^2}+V(r)\right]\psi_l(r)=E\psi_l(r). \label{CSE}
	\end{eqnarray} 
	2) The momentum space Schr\"odinger equation 
	\begin{eqnarray}
		\frac{p^2}{2m}\phi_l(p)+ \frac{1}{(2\pi)^3}\int p'^{2}dp' V_{l,l'} (p,p')\phi_{l'}(p')=E\phi_l(p). \label{MSE}
	\end{eqnarray}
	These two equations are equivalent, and we only consider the first case in this paper.
	\subsection{A brief introduction to the CSM}\label{subsec:csm}
	Before discussing the analyticity of the OPE potentials, we give a brief introduction to the CSM herein. This method was proposed by Aguilar, Balslev and Combes \cite{aguilarClassAnalyticPerturbations1971,balslevSpectralPropertiesManybody1971a} and the corresponding conclusion is called the ABC theorem. This powerful tool could directly and simultaneously get the solutions of the bound states and resonances. In this method, the resonances can be solved in the same way as the bound states. In the CSM, a simple transformation $U(\theta)$ for the radial coordinate $r$ and its conjugate momentum $k$ is introduced as:
	\begin{eqnarray}
		U(\theta)r=re^{i\theta},\qquad U(\theta)k=ke^{-i\theta}. \label{eq:rktrans}
	\end{eqnarray} 
	Then the radial Schr\"odinger equation is transformed as
	\begin{eqnarray}
		&&\Big\{\frac{1}{2m}\Big[-\frac{d^2}{dr^2}+\frac{l(l+1)}{r^2}\Big]e^{-2i\theta}+V(re^{i\theta})\Big\}\psi_l^\theta(r)\nonumber\\
		&&=E(\theta)\psi_l^\theta(r). \label{eq:SECSM}
	\end{eqnarray} 
	
\begin{figure}[htbp]
		\includegraphics[width=210pt]{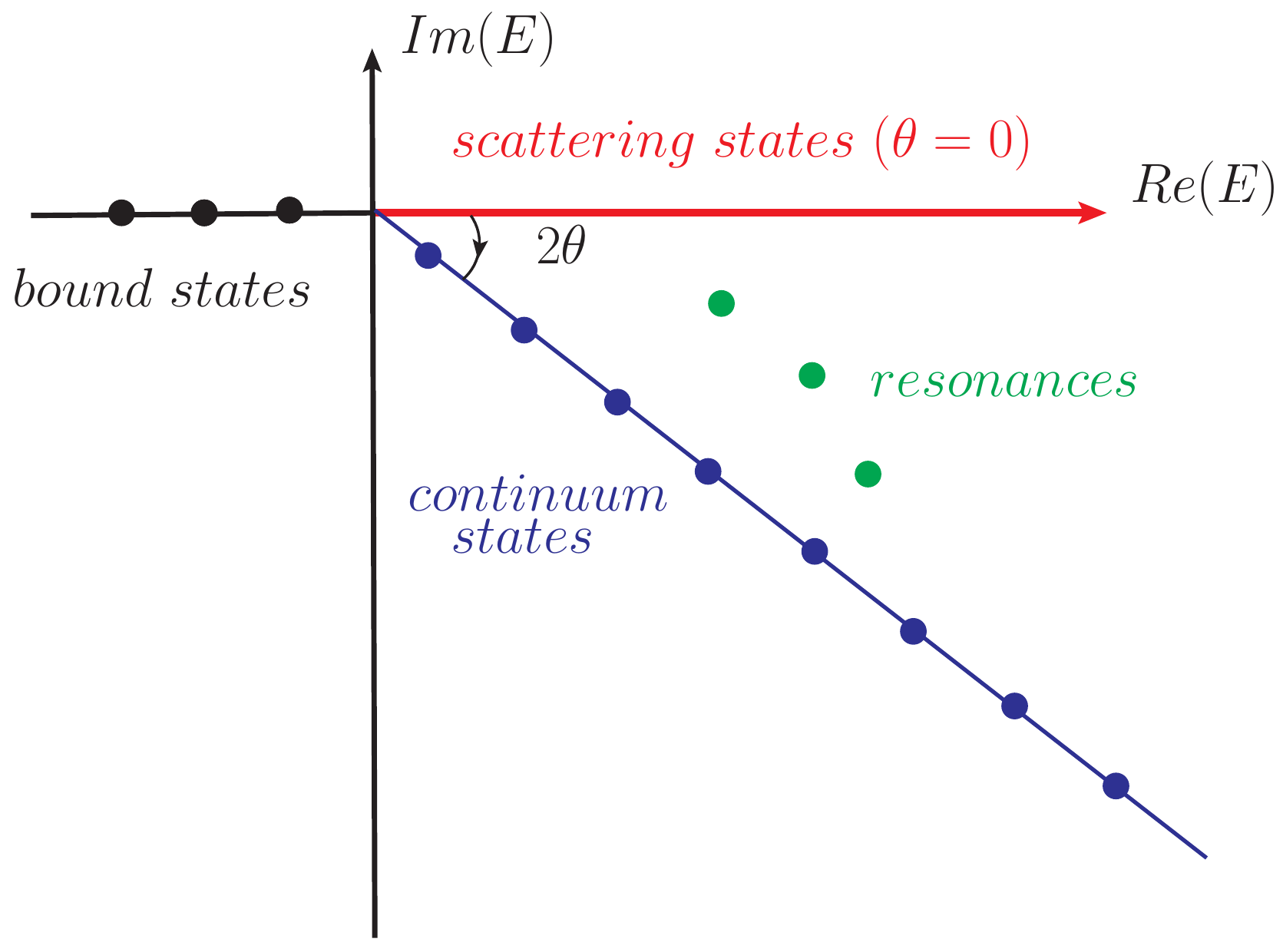}
		\caption{The eigenvalue distribution of the complex scaled Schr\"odinger equation for two-body systems. }\label{fig: CSM plot}
	\end{figure}

	As explained in the ABC theorem, one could get resonances after making a rotation on the momentum $k$. If the rotation angle $\theta$ is large enough, the resonance pole will cross the branch cut into the first Riemann sheet, as shown in Fig. \ref{fig: CSM plot}. The details can be seen in Ref. \cite{aoyamaComplexScalingMethod2006,hoMethodComplexCoordinate1983}.
	
	\subsection{Analyticity of the OPE potentials for the $DD^*$ system}\label{subsec:anlyticity Tcc}
	
	In fact, the authors of Ref. \cite{ohkodaExoticMesonsDouble2012b} used the CSM to discuss the double-heavy tetraquark system. They took an instantaneous approximation and ignored the $q_0$, which is the $0$-th component of the transferred momentum. However, this treatment may change the analyticity of the OPE potential. In this framework, we retain the $q_0$ and get some different features from the situation in the general dimeson systems. We will consider the $BB^*$ and $DD^{*}$ cases to interpret these differences. The OPE potentials of the process $\bar{B}\bar{B}^*\to \bar{B}^*\bar{B}$ and $DD^*\to D^*D$ have the same form as follows
	\begin{eqnarray}
		V_\pi=\frac{\textit{g}^2}{2f_\pi^2}\frac{(\boldsymbol{\epsilon}^*\cdot\boldsymbol{q})(\boldsymbol{\epsilon}\cdot\boldsymbol{q})}{q^2-m_\pi^2}\boldsymbol{\tau_1}\cdot\boldsymbol{\tau_2}, \label{eq:OPE}
	\end{eqnarray}
	where the $\boldsymbol{\epsilon}$ is the polarization vector of the $\bar{B}^*$ or $D^*$ meson, the $\boldsymbol{\tau_1}$ and $\boldsymbol{\tau_2}$ are the isospin operators of the $\bar{B}^{(*)}$ or $D^{(*)}$ meson. The denominator above gives $q^2-m_\pi^2=-(\boldsymbol{q}^2+m_\pi^2-q_0^2)$, where the $q_0$ is approximately equal to $m_{B^*}-m_B$ or $m_{D^*}-m_D$.
	\begin{figure}[htbp]
		\subfigure[$\bar{B}\bar{B}^{*}$ poles]{ \label{fig: Tbb poles}
			\includegraphics[width=150pt]{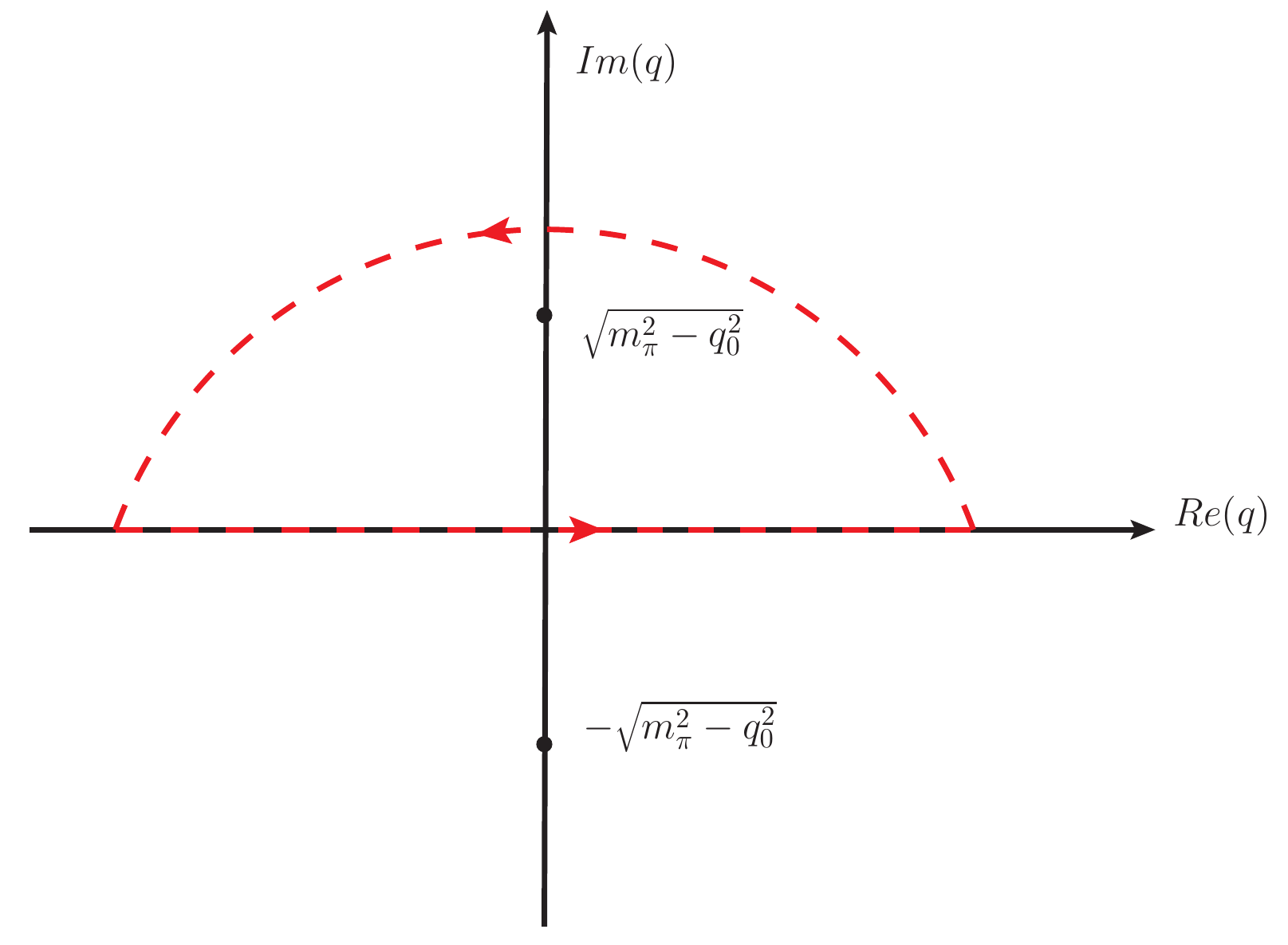}}\hspace{40pt}
		\subfigure[$DD^{*}$ PV]{ \label{fig: Tcc PV}
			\includegraphics[width=150pt]{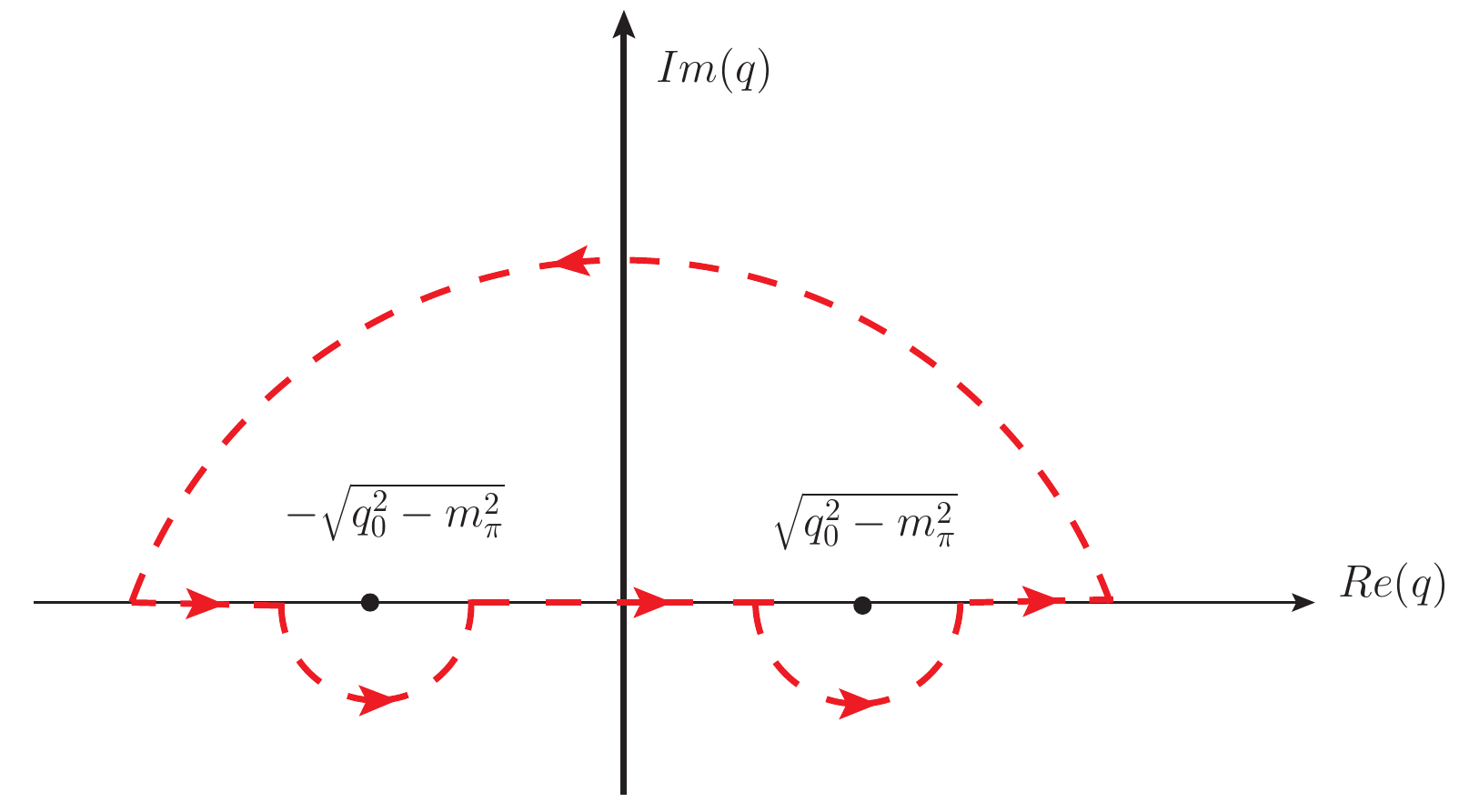}}\\
		\subfigure[$DD^{*}$ FP]{ \label{fig: Tcc FP}
			\includegraphics[width=150pt]{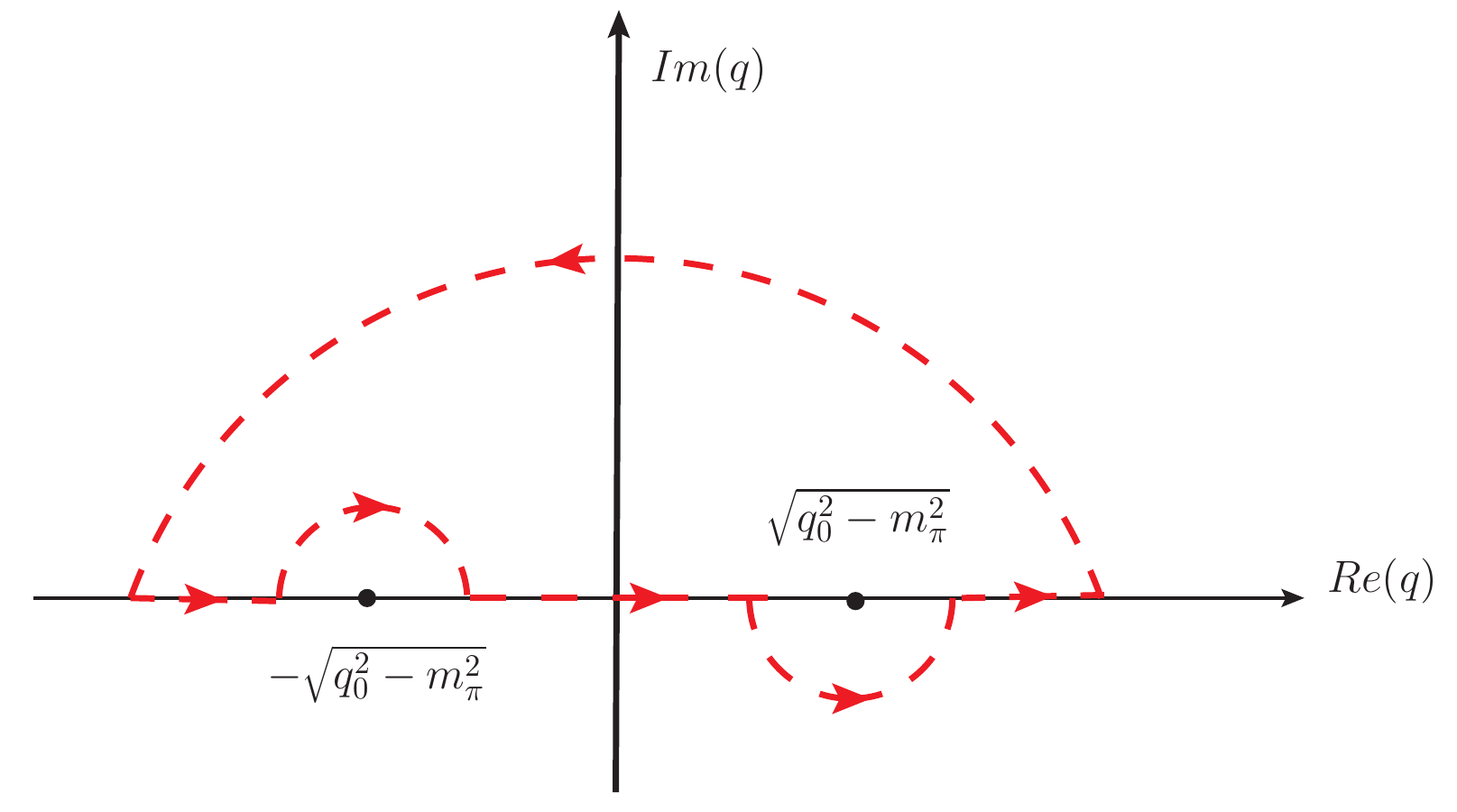}}\hspace{40pt}
		\subfigure[$DD^{*}$ CSM]{ \label{fig: Tcc CSM}
			\includegraphics[width=150pt]{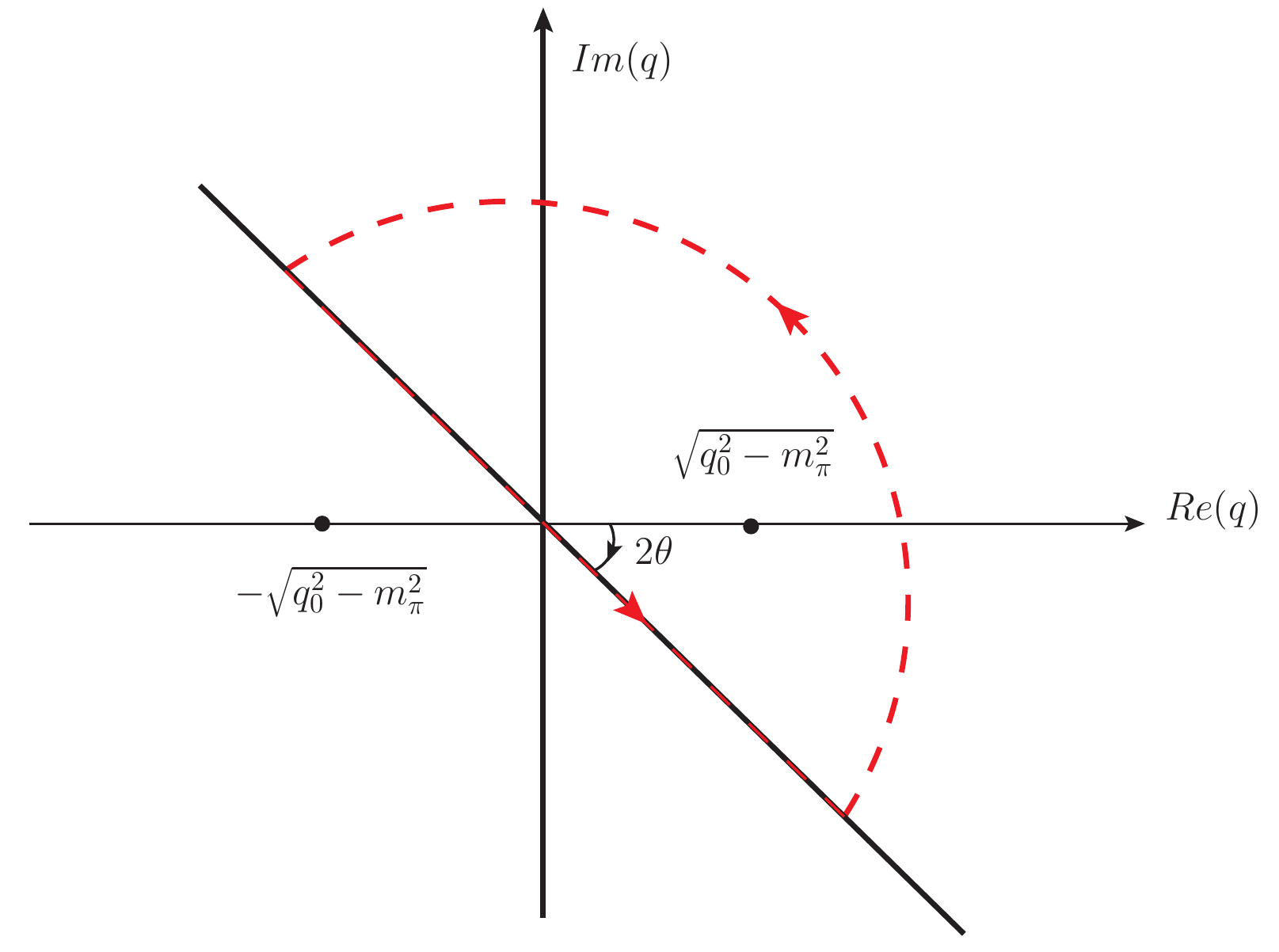}}
		\caption{For the $\bar{B}\bar{B}^{*}/DD^{*}$ systems, we show the poles of the OPE potential in the momentum space, and the corresponding contour integral (red dashed lines). We plot (a) the $\bar{B}\bar{B}^{*}$ case (b) the $DD^{*}$ case in the Cauchy principal (PV) value scheme (c) the $DD^{*}$ case in the Feynman prescription (FP) scheme (d) the $DD^{*}$ case in the complex scaling method.}
	\end{figure}
	
	For the $\bar{B}\bar{B}^*$ case, we have $q_0\approx m_{B^*}-m_B<m_\pi$. Therefore, the poles are located on the imaginary axis of the transferred momentum $|\vec{q}|$. When making the Fourier transformation to get the potential $V_\pi(r)$ in coordinate space, we need to take a contour integral in the upper half of the transferred momentum plane. The pole inside the contour will give a contribution to this integral, as illustrated in Fig. \ref{fig: Tbb poles}. In addition, this situation can also occur in the one-$\sigma/\eta/\rho/\omega$-exchange potentials for the $\bar{B}\bar{B}^*$ and $DD^*$ systems. 
	
	However, the OPE potential for the $DD^{*}$ case is different. We take the process $D^{0} D^{*+}\to D^{*+}D^0$ as an example. Since the $q_0\approx m_{D^{*+}}-m_{D^{0}}>m_{\pi^{+}}$, one could get a denominator $-(\boldsymbol{q}^2-m_{eff}^2)$, where the shorthand $m_{eff}=\sqrt{q_0^2-m_{\pi^{+}}^2}$. Obviously, the poles are located on the real transferred momentum axis, and the ambiguity emerges. In this situation, different contour integral schemes may lead to different results. One of the schemes is to take the Cauchy principal (PV) value. The OPE potential in the PV scheme is proportional to 
	{\small\begin{eqnarray}
			P\left(\frac{1}{\boldsymbol{p}^2-m_{eff}^2}\right)=\frac{1}{2}\left(\frac{1}{\boldsymbol{p}^2-m_{eff}^2+i\epsilon}+\frac{1}{\boldsymbol{p}^2-m_{eff}^2-i\epsilon}\right).\nonumber
	\end{eqnarray}}
	The contour integral is illustrated in Fig. \ref{fig: Tcc PV}. However, we will adopt another scheme---contour integral under the Feynman prescription (FP), see Fig. \ref{fig: Tcc FP}. The corresponding OPE potential is proportional to
	\begin{eqnarray}
		\frac{1}{\boldsymbol{p}^2-m_{eff}^2-i\epsilon}.\nonumber
	\end{eqnarray}
	Obviously, these two schemes will lead to different results. One can get a purely real potential in the PV scheme and a complex potential in the FP scheme. In other words, the OPE potential will contribute an imaginary part in the latter case.
	
	Then we explore the analyticity of the OPE potential in the CSM. As introduced in \ref{subsec:csm}, the CSM will rotate the propagator momentum $k$ with the angle $\theta$, and the poles of the transferred momentum become $\pm e^{i\theta}\sqrt{q_0^2-m_\pi^2}$. The pole on the positive real axis can be naturally moved into the contour region under this operation, while the other one is moved outside the region, as illustrated in Fig. \ref{fig: Tcc CSM}. It can be seen that the CSM coincides with the FP scheme shown in Fig.\ref{fig: Tcc FP}. In other words, they both choose the pole on the positive real axis, which leads to an extra imaginary potential. 
	
	In addition, when using the momentum space Schr\"odinger equation \eqref{CSE} to handle the OPE case for the $DD^{*}$ system, we will reach a pole located on the real axis of momentum. This pole will lead to the divergence in numerical calculations. In fact, the CSM could also handle this problem. Similar to the situation in the Fourier transformation, the poles will be moved from the real axis to the complex region of the momentum plane under the complex scaling operation. Then the integral along the real momentum axis will bypass the pole, and the divergence disappears.
	
	\subsection{Three-body intermediate coupled channel}
	
	As interpreted in the previous subsection, the additional imaginary contribution in the $DD^{*}$ case is from the decay process $D^{*}\to D\pi$. Therefore, the three-body coupled channel effect of $DD\pi$ cannot be neglected. In Subsec. \ref{subsec:anlyticity Tcc}, we use an approximate expression $q_0\approx m_{D^*}-m_D$ for the $DD^{*}$ case. However, this approximation is kind of rough since the $q_0$ is quite close to the $m_{\pi}$. Therefore, we should make a careful discussion on the $q_0$ herein. For simplicity, we take the $D^{0}D^{*+}\to D^{0}D^{*+}$ case as an example. 
	
	\begin{figure}[htbp]
		\includegraphics[width=220pt]{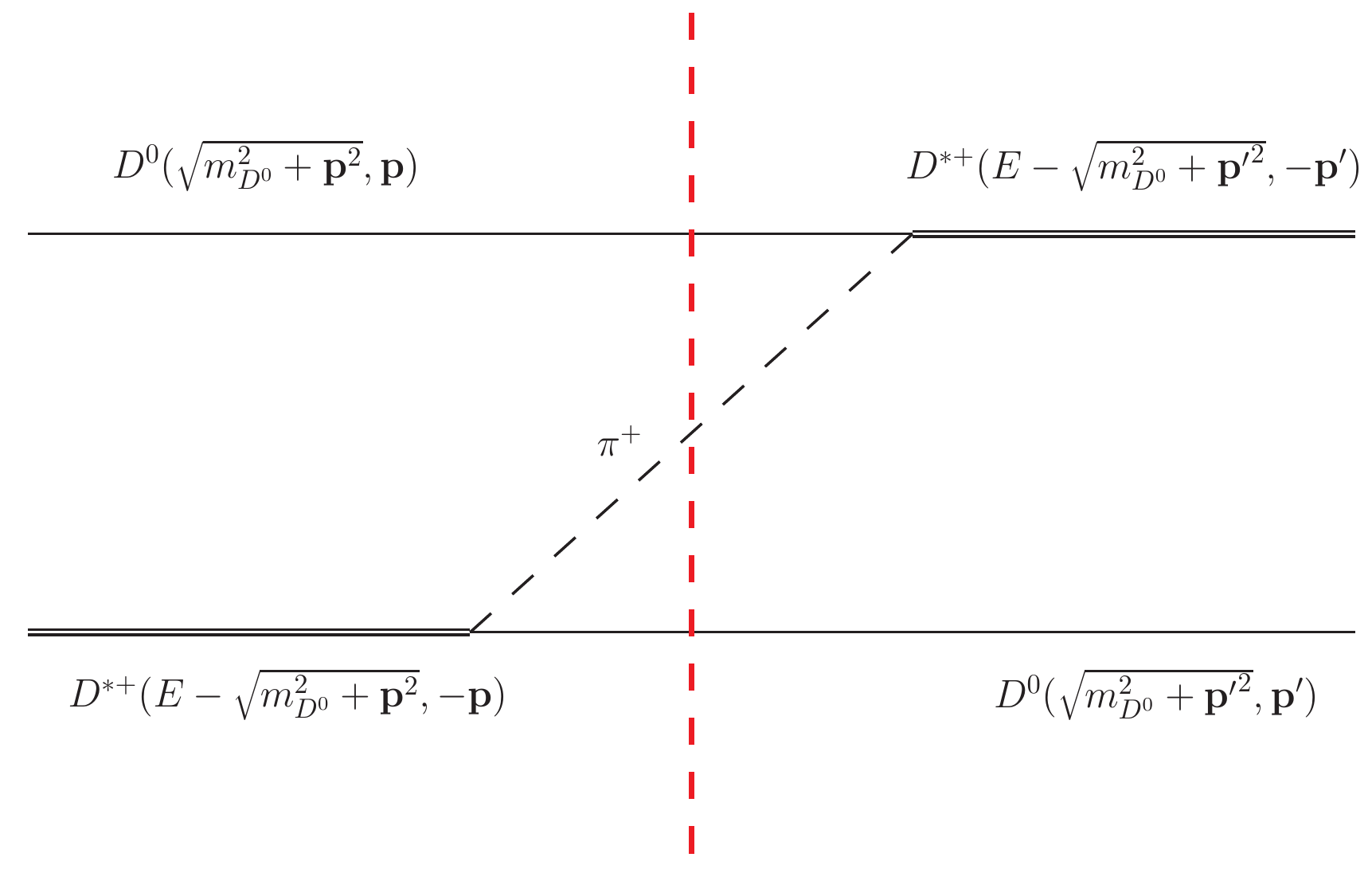}
		\caption{Three-body intermediate diagram in the process $D^0D^{*+}\to D^0D^{*+}$. The total energy of the $D^{0}D^{*+}$ is $E$, and the $D^0$ is assumed to be on shell.}\label{fig: FD Tcc}
	\end{figure}
	
	As illustrated in Fig. \ref{fig: FD Tcc}, we denote the total energy of the $D^{0}D^{*+}$ as $E$ and assume the $D^0$ meson to be on shell. Then the $q_0=E-\sqrt{m_{D^0}^2+\boldsymbol{p}^2}-\sqrt{m_{D^0}^2+\boldsymbol{p}'^{2}}$. Obviously, this diagram contains a three-body effect---the intermediate coupled channel $D^0D^0\pi^+$. When we make a non-relativistic approximation $q_0=E-2m_{D^0}-\boldsymbol{p}^2/2m_{D^0}-\boldsymbol{p}'^{2}/2m_{D^0}$, one can get a similar propagator as shown in Ref. \cite{duCoupledchannelApproachIncluding2022}. They use the time-ordered perturbation theory with the non-relativistic propagator. By contrast, we only make a non-relativistic approximation for the charmed meson $D/D^{*}$ propagator, and keep a relativistic form for the $\pi$ meson propagator. 
	
	In fact, we will neglect the kinetic energy terms $\boldsymbol{p}^2/2m_{D^0}$ and $\boldsymbol{p}'^2/2m_{D^0}$ of the charmed mesons, which are much less than the $q_0-m_\pi$. Then we make an energy shift $E\to E+m_{D^{0}}+m_{D^{*+}}$, and the potential for the $D^0D^{*+}\to D^0D^{*+}$ process gives
	\begin{eqnarray}
		&&V_{\pi}^{D^0D^{*+}\to D^0D^{*+}}(\boldsymbol{q})=-2\frac{g^2}{2f_\pi^2}\frac{(\boldsymbol{\epsilon}^*\cdot\boldsymbol{q})(\boldsymbol{\epsilon}\cdot\boldsymbol{q})}{\boldsymbol{q}^2+m_{\pi^+}^2-q_0^2},\nonumber\\ &&q_0=E+m_{D^{*+}}-m_{D^0}. \label{eq:3body}
	\end{eqnarray}
	Obviously, when solving the Schr\"odinger equation in coordinate space, one cannot get the solution as usual since the energy $E$ is unknown. However, we can still get a high precision solution by iterating the Schr\"odinger equation.

	\section{Potentials}\label{sec: potentials}
	
	The effective Lagrangians to describe the interactions for the $DD^*$ system are constructed in terms of the heavy quark symmetry and chiral symmetry. The concrete expressions of the OBE Lagrangians can be seen from Ref. \cite{liCoupledchannelAnalysisPossible2013}. 
	
	The other OPE potentials in momentum space are 
	\begin{eqnarray}
		&&V_{\pi}^{D^+D^{*0}\to D^+D^{*0}}(\boldsymbol{q})=-2\frac{g^2}{2f_\pi^2}\frac{(\boldsymbol{\epsilon}^*\cdot\boldsymbol{q})(\boldsymbol{\epsilon}\cdot\boldsymbol{q})}{\boldsymbol{q}^2+m_{\pi^{+}}^2-q_0^2},\ \text{with}\nonumber\\ &&q_0=E+m_{D^0}+m_{D^{*+}}-2m_{D^+},\ \text{and}\label{eq:3body other1}\\
		&&V_{\pi}^{D^0D^{*+}\to D^+D^{*0}}(\boldsymbol{q})=\frac{g^2}{2f_\pi^2}\frac{(\boldsymbol{\epsilon}^*\cdot\boldsymbol{q})(\boldsymbol{\epsilon}\cdot\boldsymbol{q})}{\boldsymbol{q}^2+m_{\pi^{0}}^2-q_0^2},\ \text{with} \nonumber\\ &&q_0=E+m_{D^{*+}}-m_{D^+}. \label{eq:3body other2}
	\end{eqnarray}
	We add a monopole form factor at each vertex 
	\begin{eqnarray}
		F(q)=\frac{\Lambda^2-m_\pi^2}{\Lambda^2-q^2},\label{eq:form factor}
	\end{eqnarray} 
	where the $q^2=q_0^2-\boldsymbol{q}^2$, and the $\Lambda$ is a cutoff parameter. After the Fourier transformation
	\begin{eqnarray}
		V(r)=\frac{1}{(2\pi)^3}\int d\boldsymbol{q}^3e^{-i\boldsymbol{q}\cdot\boldsymbol{r}}V(\boldsymbol{q})F^2(q),\label{eq:FT}
	\end{eqnarray} 
	we can get the coordinate space potentials
	\begin{widetext}
		\begin{eqnarray}
			V_{\pi}^{D^0D^{*+}\to D^0D^{*+}}(r)=2\frac{g^2}{2f_\pi^2}[S(\boldsymbol{\epsilon}_3^\dagger,\boldsymbol{\epsilon}_2)Y_3(\Lambda,q_0,m_{\pi^+},r)+T(\boldsymbol{\epsilon}_3^\dagger,\boldsymbol{\epsilon}_2)H_3(\Lambda,q_0,m_{\pi^+},r)],\label{eq:OPEP r1}\\
			V_{\pi}^{D^+D^{*0}\to D^+D^{*0}}(r)=2\frac{g^2}{2f_\pi^2}[S(\boldsymbol{\epsilon}_3^\dagger,\boldsymbol{\epsilon}_2)Y_3(\Lambda,q_0,m_{\pi^+},r)+T(\boldsymbol{\epsilon}_3^\dagger,\boldsymbol{\epsilon}_2)H_3(\Lambda,q_0,m_{\pi^+},r)],\label{eq:OPEP r2}\\ 
			V_{\pi}^{D^0D^{*+}\to D^+D^{*0}}(r)=-\frac{g^2}{2f_\pi^2}[S(\boldsymbol{\epsilon}_3^\dagger,\boldsymbol{\epsilon}_2)Y_3(\Lambda,q_0,m_{\pi^0},r)+T(\boldsymbol{\epsilon}_3^\dagger,\boldsymbol{\epsilon}_2)H_3(\Lambda,q_0,m_{\pi^0},r)],\label{eq:OPEP r3}
		\end{eqnarray}
	\end{widetext}
	where the $q_0$s of Eqs. (\ref{eq:OPEP r1}--\ref{eq:OPEP r3}) are listed in the Eqs. (\ref{eq:3body}--\ref{eq:3body other2}) respectively. In the above equations, $S(\mathbf{a},\mathbf{b})=\mathbf{a}\cdot\mathbf{b}$ and $T(\mathbf{a},\mathbf{b})=3(\mathbf{a}\cdot\mathbf{r})(\mathbf{b}\cdot\mathbf{r})/r^2-\mathbf{a}\cdot\mathbf{b}$. The matrix elements are given in Table \ref{tab: matrix element}. The $Y_3$, $H_3$ functions and relevant $Y$, $Z$, $H$ functions are defined as 
	\begin{eqnarray}
		&&Y(x)=\frac{e^{-x}}{x},\quad H(x)=(1+\frac{3}{x}+\frac{3}{x^2})Y(x),\nonumber\\
		&&Y_0(\Lambda,q_0,m,r)=\frac{u}{4\pi}[Y(ur)-\frac{\chi}{u}Y(\chi r)-\frac{\beta^2}{2\chi u}e^{-\chi r}],\nonumber\\
		&&Y_3(\Lambda,q_0,m,r)=\frac{u^3}{12\pi}[Y(ur)-\frac{\chi}{u}Y(\chi r)-\frac{\beta^2\chi}{2u^3}e^{-\chi r}],\nonumber\\
		&&H_3(\Lambda,q_0,m,r)=\frac{u^3}{12\pi}[H(ur)-(\frac{\chi}{u})^3 H(\chi r)\nonumber\\
		&&\qquad\qquad\qquad-\frac{\beta^2}{2\chi u}\frac{\chi^2}{u^2}Y(\chi r)-\frac{\beta^2}{2\chi u}\frac{\chi^2}{u^2}e^{-\chi r}],\label{eq:YZH}
	\end{eqnarray}
	where the
	\begin{eqnarray}
		&&u=\text{Sign}\big[\text{Re}\big(e^{i\theta}\sqrt{m^2-q_0^2}\big)\big]\sqrt{m^2-q_0^2}, \nonumber\\
		&&\beta=\sqrt{\Lambda^2-m^2},\quad \chi=\sqrt{\Lambda^2-q_0^2}.
	\end{eqnarray}
	The factor $\text{Sign}[\text{Re}(e^{i\theta}\sqrt{m^2-q_0^2})]$ can be deduced from the $DD^*$ potentials in Subsec. \ref{subsec:anlyticity Tcc}. The CSM rotation angle $\theta$ is in the region $0\le\theta\le\pi/2$. We take the $Y(ur)$ function as an example to discuss the behavior of the potentials. One can get
	$$Y(ur)=
	\left\{\begin{array}{lc}
		e^{-|u|r}/(|u|r) & m^2-q_0^2\ge0,\\
		e^{-i|u|r}(-i|u|r)&m^2-q_0^2<0.\\
	\end{array}
	\right.$$
	Obviously, the $Y(ur)$ can contribute an imaginary part to the OPE potential when $m^2-q_0^2<0$. 
	
	\begin{table}[htbp]
		\caption{The matrix elements of the operators $S(\boldsymbol{\epsilon}_3^\dagger,\boldsymbol{\epsilon}_2)$, $S(\boldsymbol{\epsilon}_4^\dagger,\boldsymbol{\epsilon}_2)$ and $T(\boldsymbol{\epsilon}_3^\dagger,\boldsymbol{\epsilon}_2)$. }\label{tab: matrix element}
		\begin{tabular}{cccc}\hline\hline
			$\Delta$&$S(\boldsymbol{\epsilon}_3^\dagger,\boldsymbol{\epsilon}_2)$&$S(\boldsymbol{\epsilon}_4^\dagger,\boldsymbol{\epsilon}_2)$&$T(\boldsymbol{\epsilon}_3^\dagger,\boldsymbol{\epsilon}_2)$\\
			\hline
			$\langle ^3S_1|\Delta|^3S_1\rangle$&1&1&$0$\\
			$\langle ^3D_1|\Delta|^3S_1\rangle$&0&0&$-\sqrt{2}$\\
			$\langle ^3S_1|\Delta|^3D_1\rangle$&0&0&$-\sqrt{2}$\\
			$\langle ^3D_1|\Delta|^3D_1\rangle$&1&1&$1$\\ 
			\hline\hline
		\end{tabular}
	\end{table} 
	
	We also consider the short- and medium-range one-$\sigma,\eta,\rho,\omega$-exchange potentials from the OBE interactions. The corresponding potentials are given in Eqs. (\ref{eq:OBEP eta}--\ref{eq:OBEP omega}). These potentials have no imaginary contributions due to the large exchanged meson mass, and the isospin breaking effects barely have influence on the one-$\sigma,\eta,\rho,\omega$-exchange potentials. 
	\begin{widetext}
		\begin{eqnarray}
			V_{\eta}^{D^0D^{*+}\to D^+D^{*0}}(r)=&&\frac{1}{3}\frac{g^2}{2f_\pi^2}[S(\boldsymbol{\epsilon}_3^\dagger,\boldsymbol{\epsilon}_2)Y_3(\Lambda,q_0^C,m_{\eta},r)+T(\boldsymbol{\epsilon}_3^\dagger,\boldsymbol{\epsilon}_2)H_3(\Lambda,q_0^C,m_{\eta},r)], \label{eq:OBEP eta}\\
			V_{\sigma}^{D^0D^{*+}\to D^0D^{*+}}(r)=&&V_{\sigma}^{D^+D^{*0}\to D^+D^{*0}}(r)=-g_\sigma^2 S(\boldsymbol{\epsilon}_4^\dagger,\boldsymbol{\epsilon}_2)Y_0(\Lambda,q_0^D,m_{\sigma},r),\label{eq:OBEP sigma}\\
			V_{\rho}^{D^0D^{*+}\to D^0D^{*+}}(r)=&&V_{\rho}^{D^+D^{*0}\to D^+D^{*0}}(r)=-\frac{1}{4}\beta^2 g_V^2 S(\boldsymbol{\epsilon}_4^\dagger,\boldsymbol{\epsilon}_2)Y_1(\Lambda,q_0^D,m_{\rho},r)\nonumber\\
			&&+2\lambda^2 g_V^2[2S(\boldsymbol{\epsilon}_3^\dagger,\boldsymbol{\epsilon}_2)Y_3(\Lambda,q_0^C,m_{\rho},r)-T(\boldsymbol{\epsilon}_3^\dagger,\boldsymbol{\epsilon}_2)H_3(\Lambda,q_0^C,m_{\rho},r)],\nonumber\\ 
			V_{\rho}^{D^0D^{*+}\to D^+D^{*0}}(r)=&&2\frac{1}{4}\beta^2 g_V^2 S(\boldsymbol{\epsilon}_4^\dagger,\boldsymbol{\epsilon}_2)Y_1(\Lambda,q_0^D,m_{\rho},r)-\lambda^2 g_V^2[2S(\boldsymbol{\epsilon}_3^\dagger,\boldsymbol{\epsilon}_2)Y_3(\Lambda,q_0^C,m_{\rho},r)\nonumber\\
			&&-T(\boldsymbol{\epsilon}_3^\dagger,\boldsymbol{\epsilon}_2)H_3(\Lambda,q_0^C,m_{\rho},r)],\label{eq:OBEP rho}\\ 
			V_{\omega}^{D^0D^{*+}\to D^0D^{*+}}(r)=&&V_{\omega}^{D^+D^{*0}\to D^+D^{*0}}(r)=\frac{1}{4}\beta^2 g_V^2 S(\boldsymbol{\epsilon}_4^\dagger,\boldsymbol{\epsilon}_2)Y_1(\Lambda,q_0^D,m_{\omega},r),\nonumber\\ 
			V_{\omega}^{D^0D^{*+}\to D^+D^{*0}}(r)=&&\lambda^2 g_V^2[2S(\boldsymbol{\epsilon}_3^\dagger,\boldsymbol{\epsilon}_2)Y_3(\Lambda,q_0^C,m_{\omega},r)-T(\boldsymbol{\epsilon}_3^\dagger,\boldsymbol{\epsilon}_2)H_3(\Lambda,q_0^C,m_{\omega},r)].\label{eq:OBEP omega} 
		\end{eqnarray}
	\end{widetext}
	So, we ignore the mass difference between the $D^{0}$($D^{*0}$) and $D^{+}$($D^{*+}$) herein. Then the $0-$th components of the transferred momenta have simple forms
	\begin{eqnarray}
		&&q_{0}^D=0, \quad q_{0}^C=m_{D^{*+}}-m_{D^{0}},
	\end{eqnarray}
	where the $q_{0}^D$ and $q_{0}^C$ correspond to the direct and cross diagrams in the processes $DD^*\to DD^*$.
	
	So far, we get all the OBE potentials in Eqs. (\ref{eq:OPEP r1}--\ref{eq:OPEP r3}) and Eqs. (\ref{eq:OBEP eta}--\ref{eq:OBEP omega}).  In addition, we give the adopted parameter value in the OBE potentials. The $f_\pi=132$ MeV is the pion decay constant. The coupling constants for the $\pi$ and $\sigma$ exchange are $g=0.59\pm0.07\pm0.01$ \cite{ahmedFirstMeasurement2001} and $g_\sigma=3.73/2\sqrt{6}$ \cite{liu4430MolecularState2008a} respectively. And we adopt the coupling constants related to the vector meson exchange, $g_V=5.8$, $g_\beta=0.9$ and $g_\lambda=0.56$ GeV$^{-1}$ \cite{bandoNonlinearRealizationHidden1988,isolaCharmingPenguinContributions2003}.

	\section{Numerical results}\label{sec: results}
	\subsection{The OPE potential results for the $DD^*$ system}
	
\begin{figure}[htbp]
		\includegraphics[width=240pt]{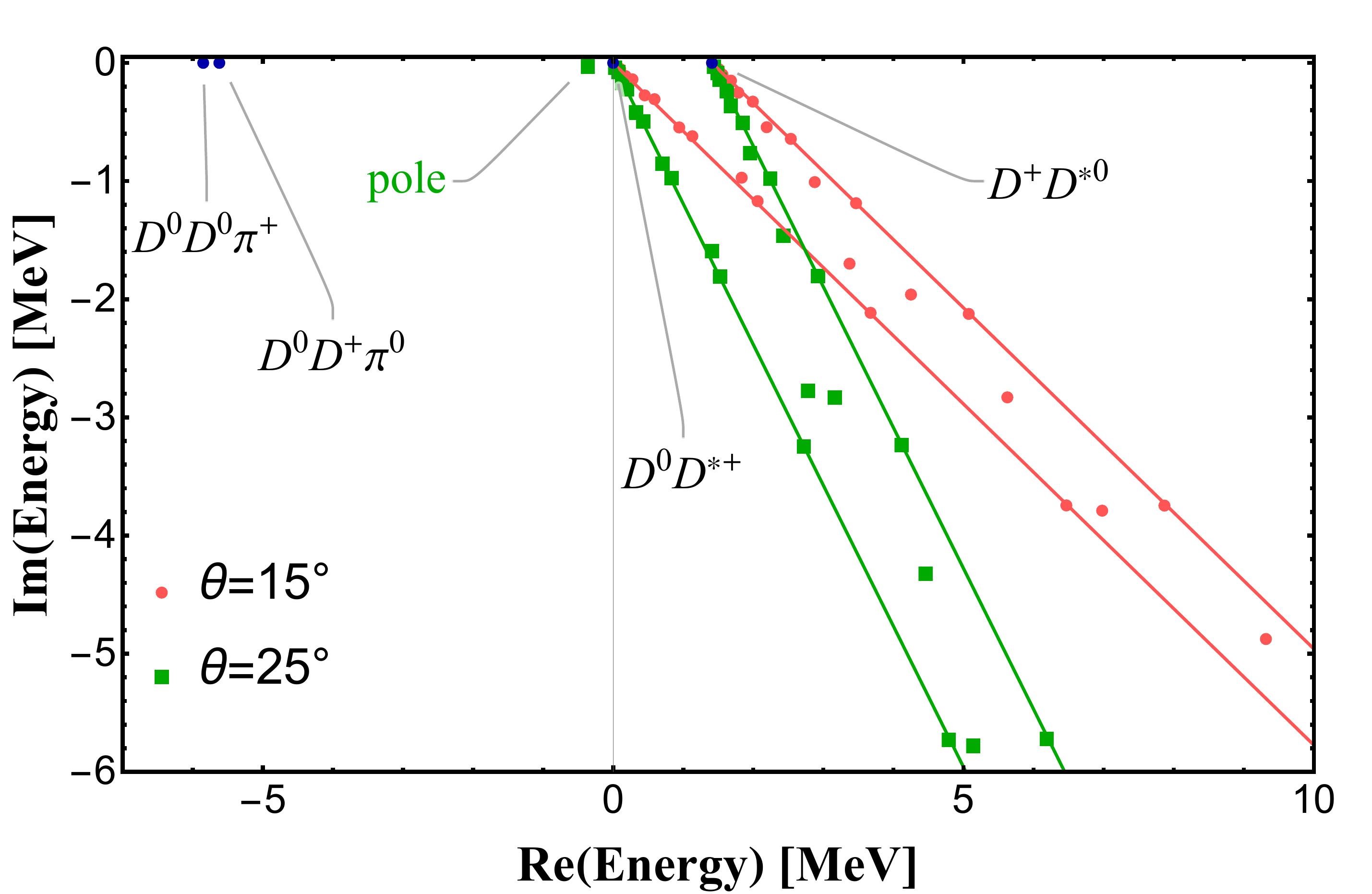}
		\caption{The eigenvalue distribution for the OPE potential case with the $\Lambda=1602$ MeV. The red (green) points (square point) and lines correspond to the situation with the complex rotation angle $\theta=15^{\circ}\ (25^{\circ})$.}\label{fig: Tcc csm opep}
	\end{figure}

	We first introduce the OPE potentials only to study the $DD^*$ system. The Gaussian expansion method (GEM) \cite{hiyamaGaussianExpansionMethod2003} is adopted to solve the complex scaled Schr\"odinger equation, see Eq. \eqref{eq:SECSM}. In this calculation, to determine the only unknown parameter $\Lambda$, we take the mass of the $T_{cc}^{+}$ as an input. Then we get the cutoff $\Lambda=1602$ MeV, and illustrate the CSM eigenvalue distribution in Fig. \ref{fig: Tcc csm opep}. We plot two lists of eigenvalues obtained from rotational angles $\theta=15^\circ,25^\circ$. Only one pole is found with the energy $E=-364-31 i$ keV relative to the threshold of $D^{0}D^{*+}$. This pole could be interpreted as a quasibound state corresponding to the $T_{cc}^{+}$.
	
\begin{figure}[htbp]
		\subfigure[Re$\big(u_{i}(r)\big)$]{ \label{fig: ur opep re}
			\includegraphics[width=210pt]{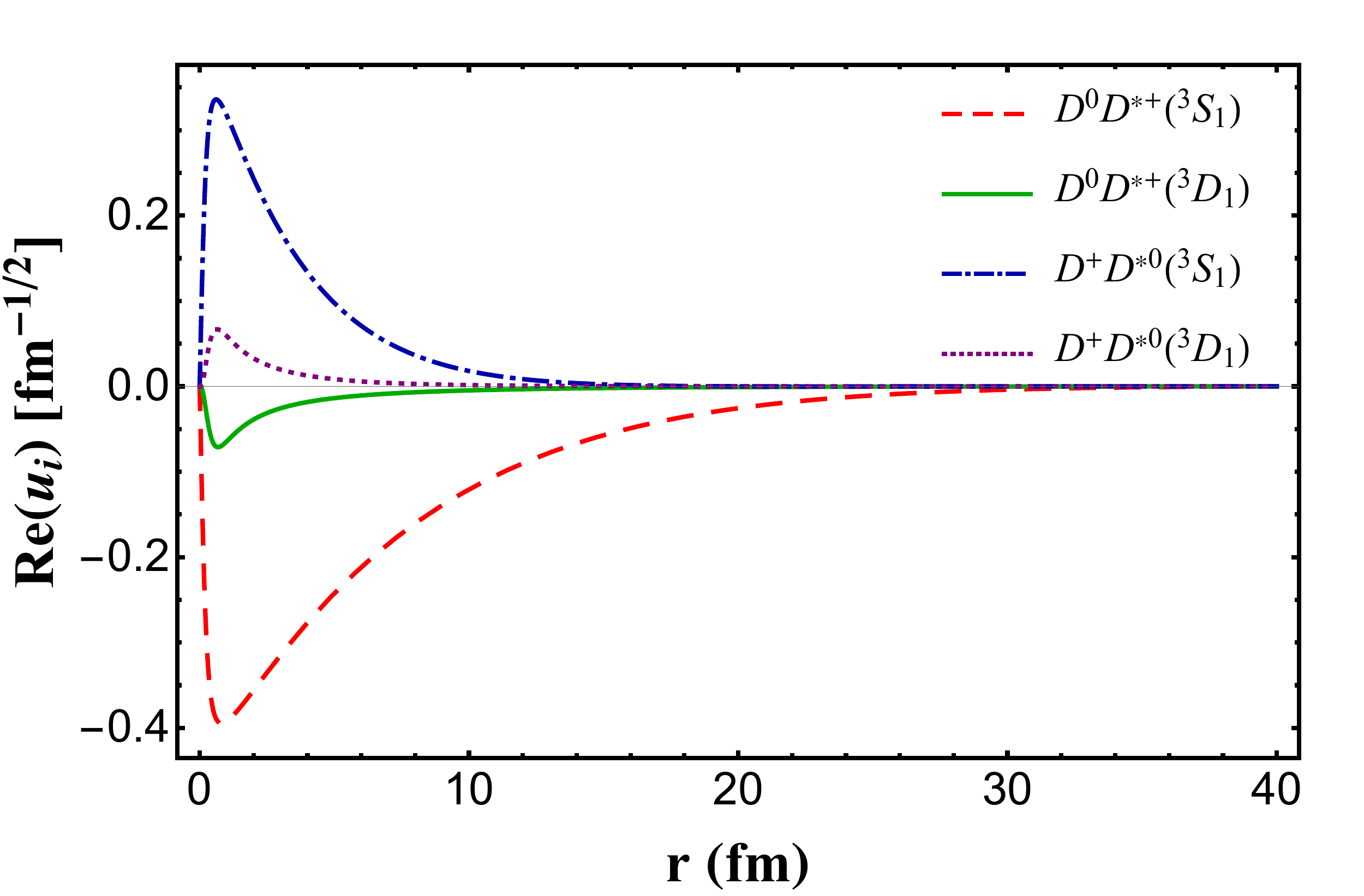}}\hspace{40pt}
		\subfigure[Im$\big(u_{i}(r)\big)$]{ \label{fig: ur opep im}
			\includegraphics[width=210pt]{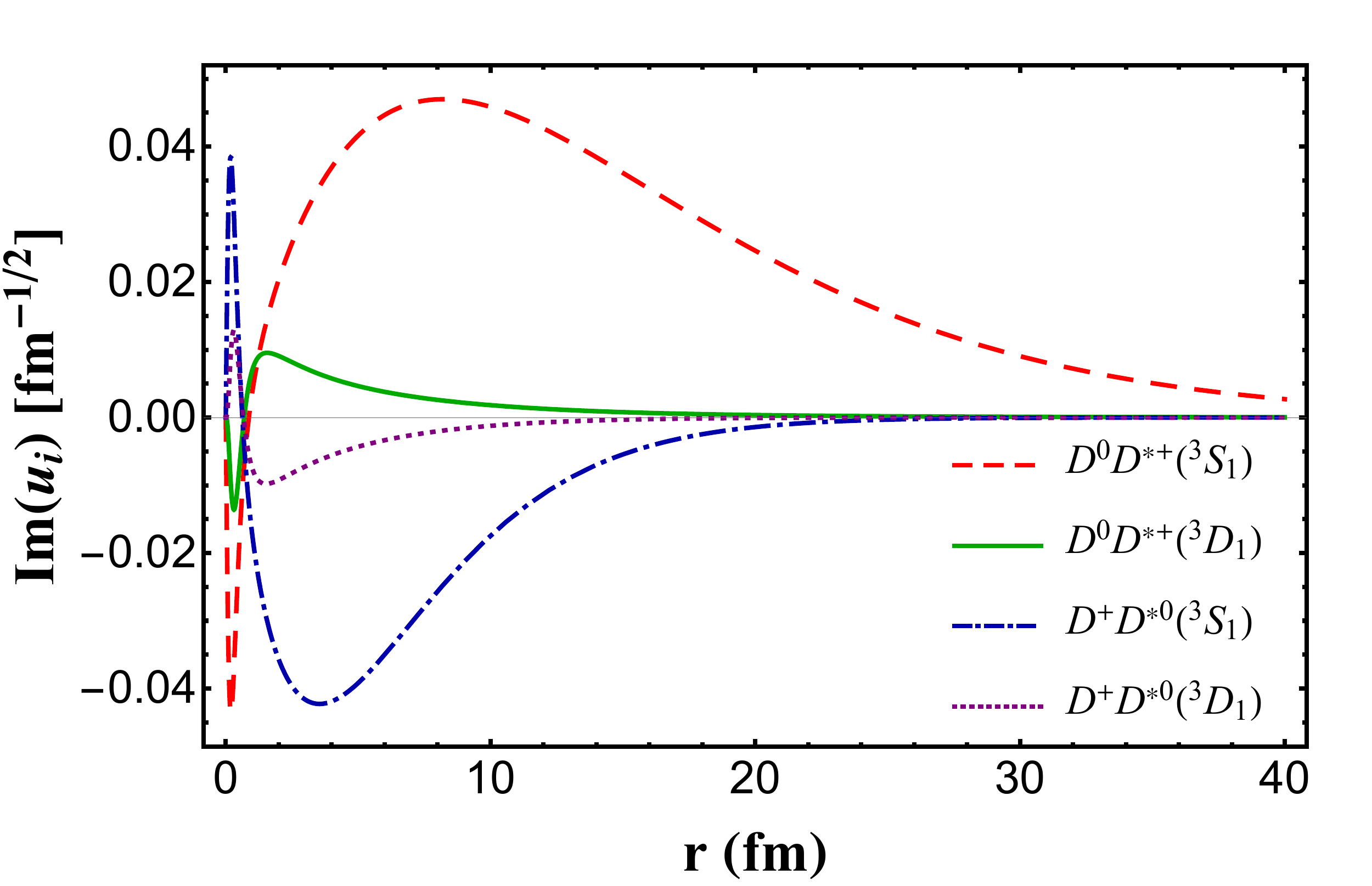}}
		\caption{The wave functions $u_i(r)(i=1,2,3,4)$ for the OPE potential case. The rotation angle $\theta=15^\circ$ and the cutoff $\Lambda=1602$ MeV. The two diagrams correspond to: (a) the real part of the $u_i(r)$ (b) the imaginary part of the $u_i(r)$.}\label{fig: ur opep}
	\end{figure}

	Considering the coupled channels $D^{0}D^{*+}$ and $D^{+}D^{*0}$ ($D^{0}D^{0}\pi^+$ and $D^{0}D^{+}\pi^0$), we could find the branch cuts with their branch point at the two-body (three-body) thresholds. Obviously, the quasibound state pole in Fig. \ref{fig: Tcc csm opep} is located on the first Riemann sheets (physical sheets) corresponding to the $D^{0}D^{*+}$ and $D^{+}D^{*0}$ channels and the second Riemann sheets (unphysical sheets) corresponding to the $D^{0}D^{0}\pi^+$ and $D^{0}D^{+}\pi^0$ channels. In Fig. \ref{fig: ur opep}, we choose $\theta=15^\circ$ and present the real and imaginary part of its wave function. The probabilities of the channels $D^{0}D^{*+}(^3S_1,^3D_1)$ and $D^{+}D^{*0}(^3S_1,^3D_1)$ are shown in Table \ref{tab: results}. Obviously, the $S$-wave $D^{0}D^{*+}$ and $D^{+}D^{*0}$ channels dominate this quasibound state, which is similar to the result of Ref. \cite{chenPredictingAnotherDoubly2021}. However, we do not find the resonance predicted in their work.
	
	\subsection{The OBE potential results for the $DD^*$ system}
\begin{figure}[htbp]
		\includegraphics[width=220pt]{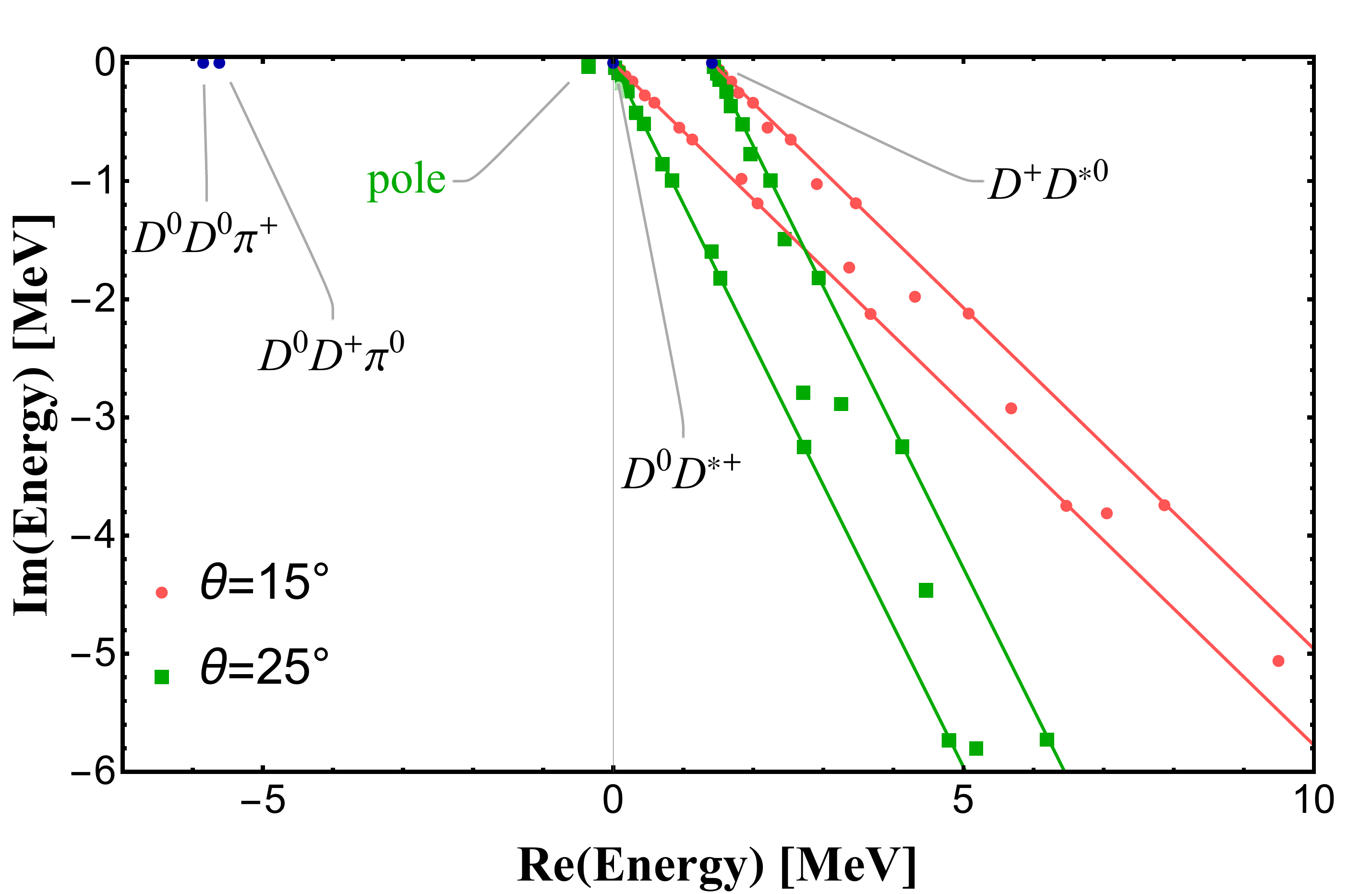}
		\caption{The eigenvalue distribution in the OBE potential case with the $\Lambda=1170$ MeV. The red (green) points (square point) and lines correspond to the situation with the complex rotation angle $\theta=15^{\circ}\ (25^{\circ})$.}\label{fig: Tcc csm obep}
	\end{figure}

\begin{figure}[htbp]
		\subfigure[Re$\big(u_{i}(r)\big)$]{ \label{fig: ur obep re}
			\includegraphics[width=210pt]{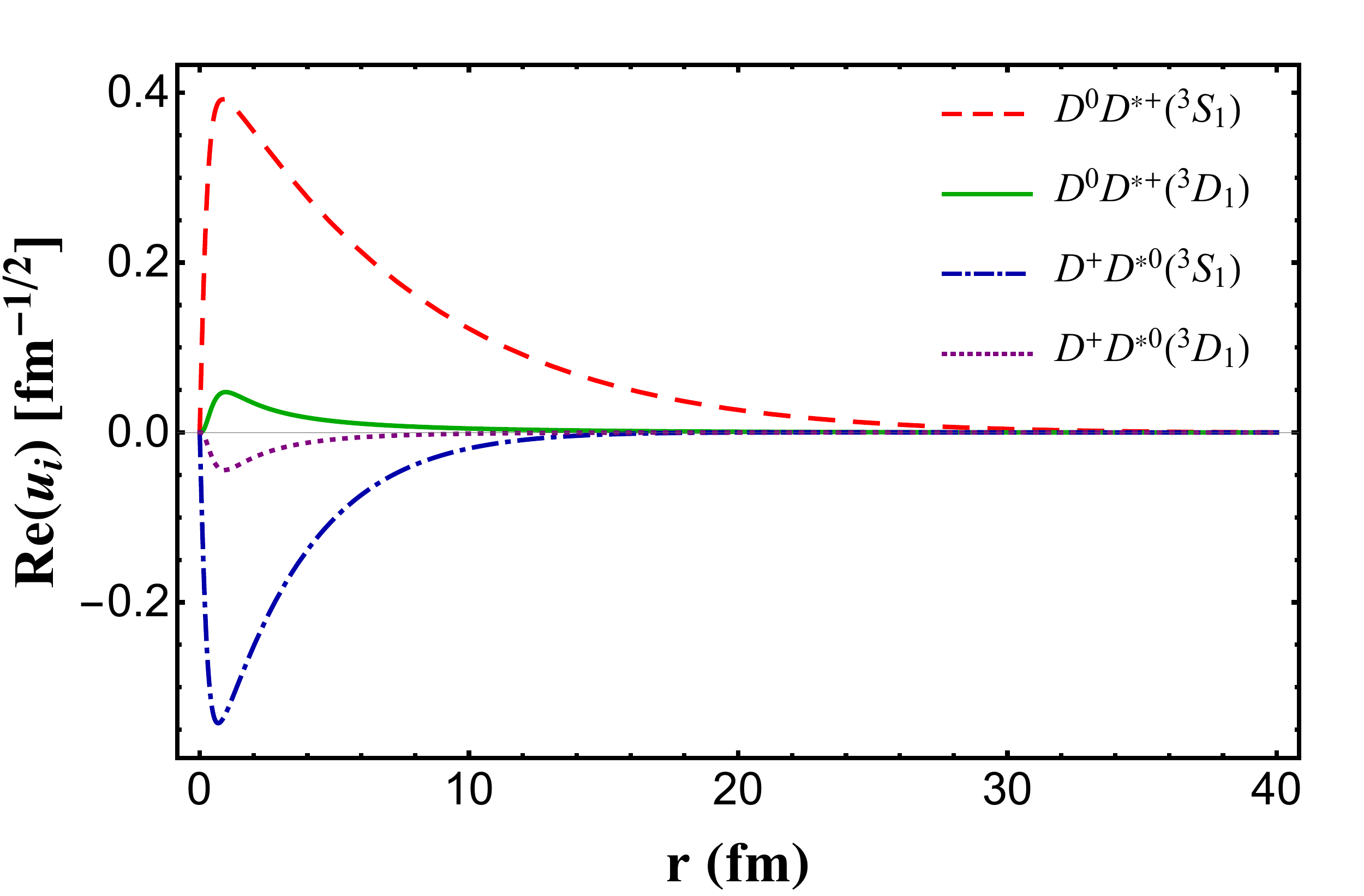}}\hspace{40pt}
		\subfigure[Im$\big(u_{i}(r)\big)$]{ \label{fig: ur obep im}
			\includegraphics[width=210pt]{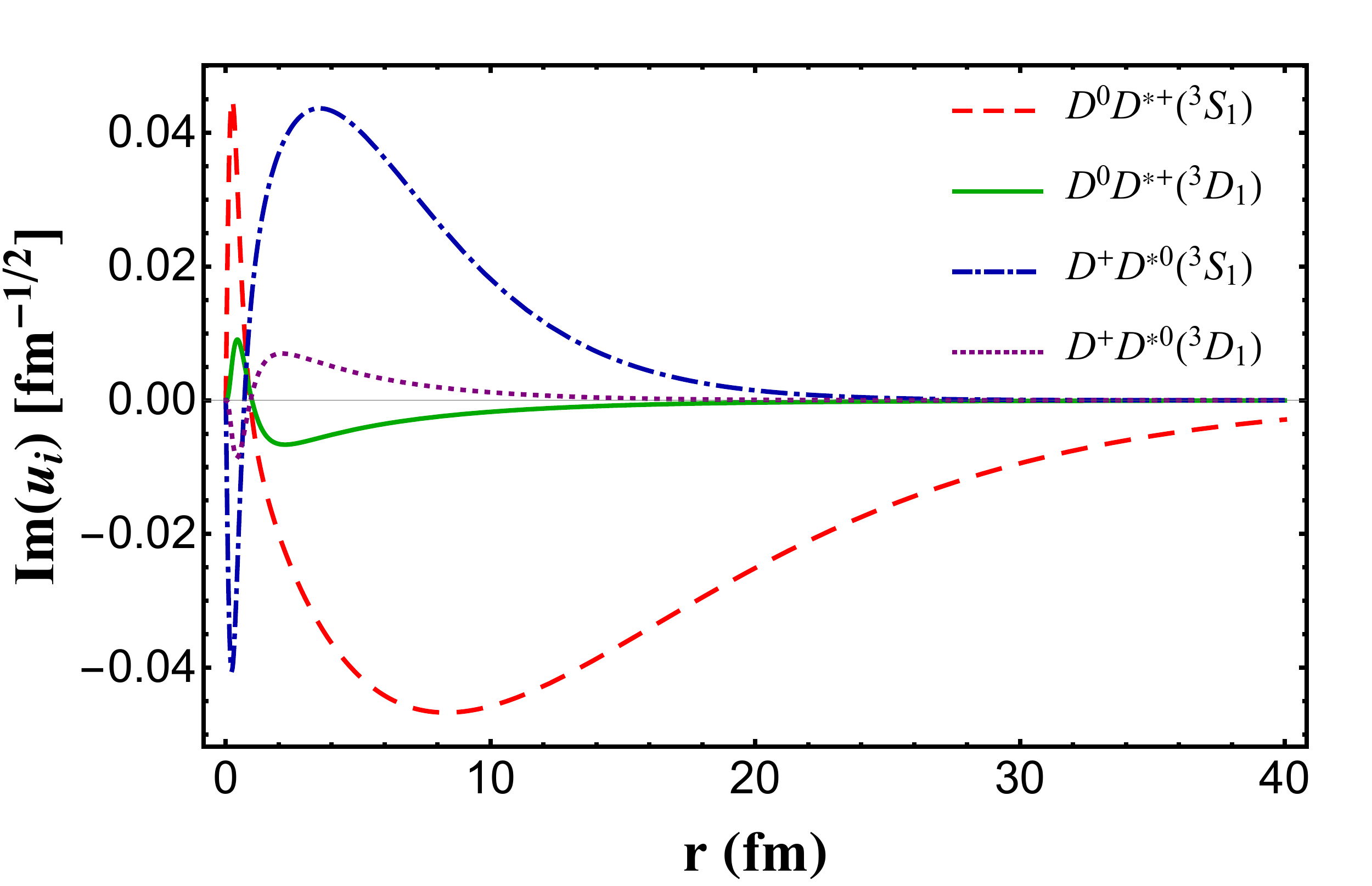}}
		\caption{The wave functions $u_i(r)(i=1,2,3,4)$ in the OBE potential case. The rotation angle $\theta=15^\circ$ and the cutoff $\Lambda=1170$ MeV. The two diagrams correspond to: (a) the real part of the $u_i(r)$ (b) the imaginary part of the $u_i(r)$.}\label{fig: ur obep}
	\end{figure}

	In this part, we further employ the OBE potential to include the short- and medium-range contribution. When taking the mass of the $T_{cc}^+$ as an input, we use the cutoff $\Lambda=1170$ MeV and get the CSM eigenvalue distribution as illustrated in Fig. \ref{fig: Tcc csm obep}. Similar to the OPE potential case, no resonance is found herein. The only pole of the quasibound state has the energy $E=-354-31 i$ keV relative to the threshold of the $D^{0}D^{*+}$. The wave functions with $\theta=15^\circ$ are also shown in Fig. \ref{fig: ur obep}. The probabilities for the four channels are also shown in Table \ref{tab: results}. One can see that the $S$-wave $D^{0}D^{*+}$ and $D^{+}D^{*0}$ channels still dominate this quasibound state. We interpret this pole as the $T_{cc}^{+}$, and its width also coincides with the experimental data nicely.
	
	Interestingly, after we take the binding energy as input, the OPE and OBE cases give quite similar results, including the width, constituent and size. When we consider the long-range behavior of the $T_{cc}^{+}$, such a similarity turns out to be reasonable. In our framework, the $T_{cc}^{+}$ is assumed to be a molecule with a very small binding energy, which has a very large size. The RMS in Table \ref{tab: results} also supports this speculation. Therefore, the long-range OPE interactions should be dominant in this system, and we should obtain similar results with these two schemes.
	
\begin{figure}[htbp]
		\includegraphics[width=200pt]{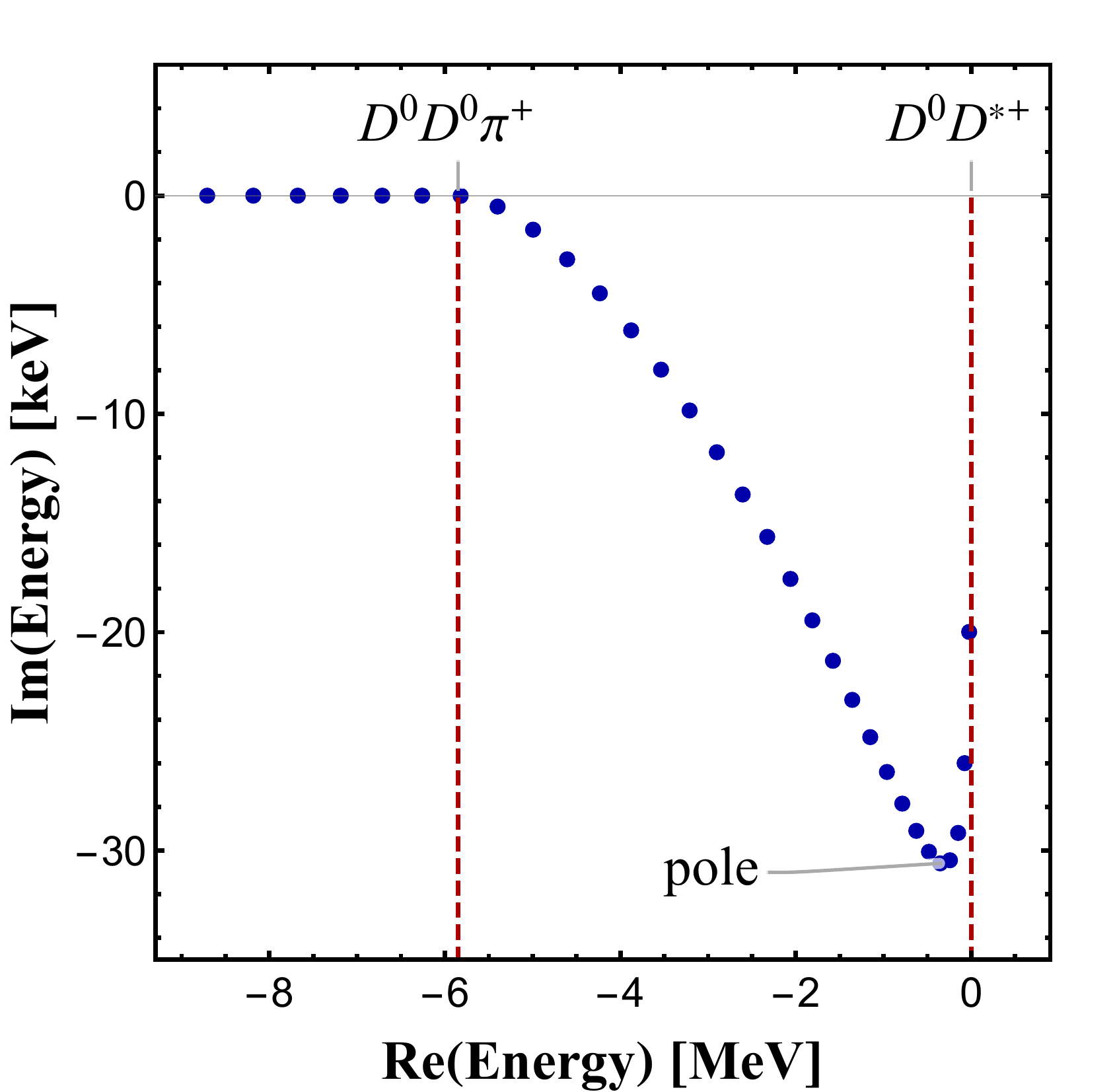}
		\caption{The dependence of the width of $DD^*$ system on the binding energy relative to the $D^{0}D^{*+}$ threshold in the OBE potential case. The width is equal to $-2$Im(Energy), and the complex rotation angle is $\theta=15^{\circ}$. The pole in this figure corresponds to the $T_{cc}^+$ with the $\Lambda=1170$ MeV.}\label{fig: Tcc energy width obep}
	\end{figure}

	In addition, we adjust the cutoff $\Lambda$ and get a list of eigenvalues to see the correlation between the binding energy and width of this pole. As presented in Fig. \ref{fig: Tcc energy width obep}, the width decreases rapidly as the binding energy becomes deeper. And the width vanishes when the energy is smaller than the thresholds of $D^0D^0\pi^+$ and $D^0D^+\pi^0$. This threshold effect is reasonable since we do not consider the electromagnetic decay process in this work. Thus, when the $DD\pi$ channel is closed, the imaginary part of the pole disappears, and the pole becomes a stable bound state.

	\begin{widetext}
	\begin{center} 
		\begin{table}[htbp]
			\renewcommand{\arraystretch}{1.8}{
				\setlength\tabcolsep{8pt}{
					\begin{tabular}{ccccccc}\hline\hline
						\centering
						System&$J^{PC}$&Potentials&$\Lambda$ (MeV)&$E$ (keV)&Probabilities&RMS (fm)\\\hline
						$DD^{*}$&$1^+$&OPE&$1602$&$-364-31i$&$(72.6,0.8,26.0,0.6)\%$&$4.9-0.1i$\\ 
						&$1^+$&OBE&$1170$&$-354-31i$&$(72.1,0.5,27.1,0.3)\%$&$5.0-0.1i$\\ 
						$[D\bar{D}^*]$&$1^{++}$&OBE&$1155$&$-111-13i$&$(92.7,0.3,6.6,0.3)\%$&$7.3-0.2i$\\ 
						\hline\hline
			\end{tabular}}}
			\caption{The numerical results for the $DD^*$ and $[D\bar{D}^{*}]$ systems with the OPE and OBE potentials, and the rotation angle is $\theta=15^{\circ}$. The $E$ is the quasibound state energy of the $DD^{*}$ ($[D\bar{D}^*]$) system relative to the threshold of $D^{0}D^{*+}$ ($D^{0}\bar{D}^{*0}$). The probabilities correspond to the channels presented in Table \ref{tab: channel}. The RMS is the root-mean-square radius in the CSM, which has been discussed in the Ref. \cite{hommaMatrixElementsPhysical1997}. Its real part is interpreted as an expectation value, and the imaginary part corresponds to a measure of the uncertainty in observation.}\label{tab: results}
		\end{table} 
	\end{center}
\end{widetext}
	
	\subsection{The OBE potential results for the $[D\bar{D}^*]$ system}
	
	We also discuss the molecule system $[D\bar{D}^*]$ with $J^{PC}=1^{++}$, where the shorthand notation $[D\bar{D}^*]$ is $\frac{1}{\sqrt{2}}(D\bar{D}^{*}-D^{*}\bar{D})$.
	Considering the isospin breaking effect, we adopt the channels $[D^{0}\bar{D}^{*0}](^3S_1,^3D_1)$ and $[D^{+}\bar{D}^{*-}](^3S_1,^3D_1)$, see Table \ref{tab: channel}. The OPE potentials of the $[D\bar{D}^*]$ are given in Eq. \eqref{eq:OPEP r3 3872}, and the other terms of the OBE potential can be found in Ref. \cite{liIsospinBreakingCoupledchannel2012}. The matrix elements $S(\boldsymbol{\epsilon}_3^\dagger,\boldsymbol{\epsilon}_2)$, $S(\boldsymbol{\epsilon}_4^\dagger,\boldsymbol{\epsilon}_2)$, $T(\boldsymbol{\epsilon}_3^\dagger,\boldsymbol{\epsilon}_2)$ can also be obtained from Table \ref{tab: matrix element}.
	
\begin{figure}[htbp]
		\includegraphics[width=220pt]{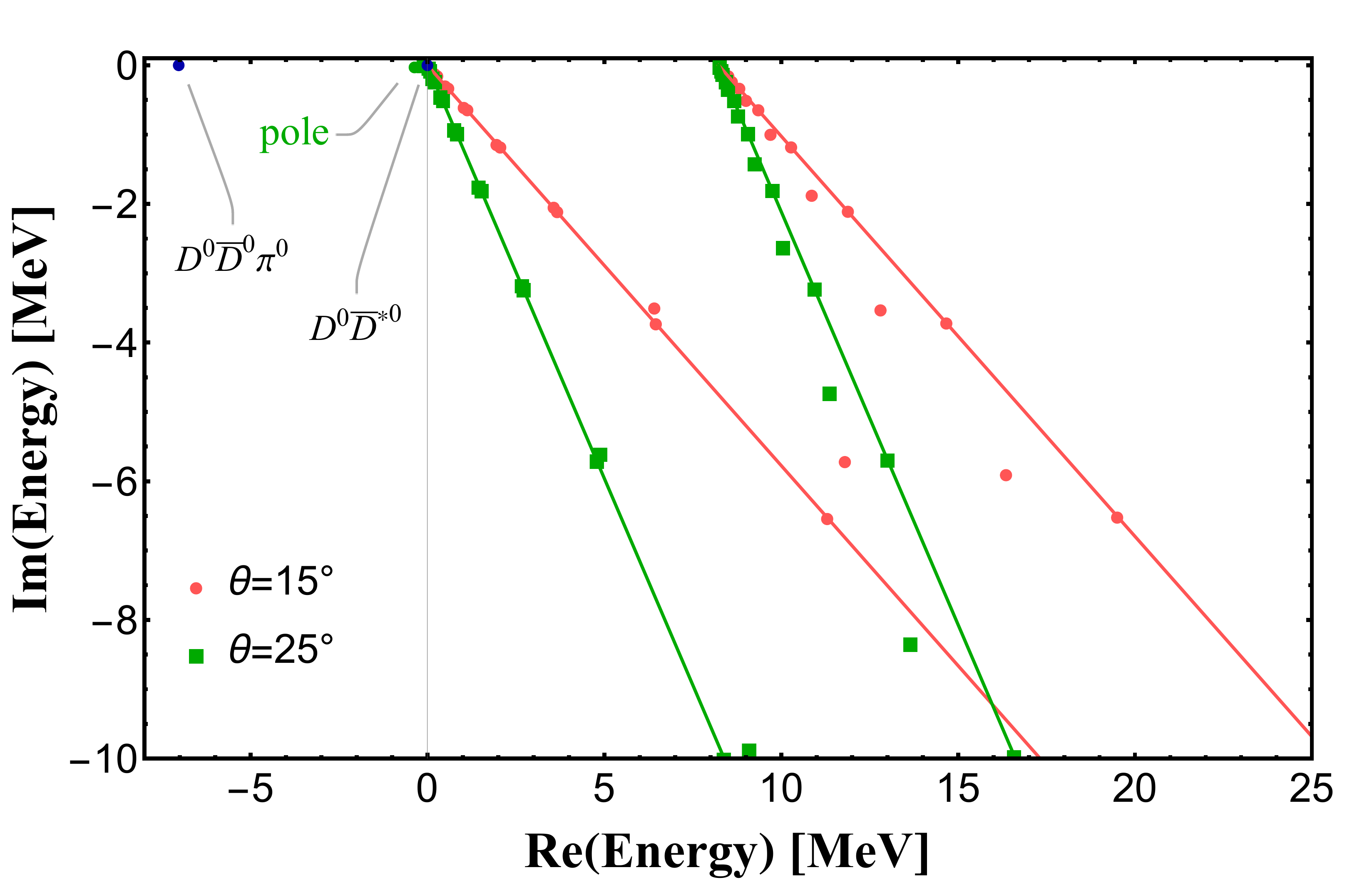}
		\caption{The eigenvalue distribution in the OBE potential case with the $\Lambda=1155$ MeV. The red (green) points (square point) and lines correspond to the case with the complex rotation angle $\theta=15^{\circ}\ (25^{\circ})$.}\label{fig: 3872 csm obep}
	\end{figure}

	In this work, we need the mass of the $X(3872)$ as input. Recently, the LHCb Collaboration studied the lineshape by a Flatt$\acute{e}$ inspired model \cite{aaijStudyLineshape38722020} and gave a new result,
	\begin{eqnarray}
		&&m_{\text{mode}}=3871.69^{+0.00+0.05}_{-0.04-0.13} \ \text{MeV/c}^2, \nonumber\\
		&&\Gamma_{\text{FWHM}}=0.22^{0.06+0.25}_{-0.08-0.17} \ \text{MeV}.\label{lhcb 3}
	\end{eqnarray} 
	One can see the mass of the $X(3872)$ is extremely close to the $D^{0}\bar{D}^{*0}$ threshold, so the accurate binding energy relative to the latter is hard to be determined. 
	We take a binding energy $-111$ keV relative to the $[D^{0}\bar{D}^{*0}]$ as input within the experimental error, and the corresponding cutoff is $\Lambda=1155$ MeV. Then we get the CSM eigenvalue distribution as illustrated in Fig. \ref{fig: 3872 csm obep}. 
	
	Similar to the $DD^{*}$ system, we only find a quasibound state with energy $E=-111-13 i$ keV. The pole is located on the first Riemann sheets (physical sheets) corresponding to the $[D^{0}\bar{D}^{*0}]$ and $[D^{+}\bar{D}^{*-}]$ channels and the second Riemann sheet (physical sheets) corresponding to the $D^{0}\bar{D}^{0}\pi^{0}$ channel. The probabilities for the $[D\bar{D}^{*}]$ channels are shown in Table \ref{tab: results}, and the $S-$wave $[D^{0}\bar{D}^{*0}]$ dominates the system.
	
	The width value is smaller than the central value in Eq. \eqref{lhcb 3} given by the LHCb Collaboration. However, our width value is still consistent with the new measurement within one standard deviation if one considers the large experimental errors. This possible difference may arise from the neglect of the other channels such as the hidden-charm channels $J/\psi\rho$, $\eta_c\omega$ and the electromagnetic channel $D^{0}\bar{D}^{0}\gamma$, which may provide a considerable contribution.
	
\begin{figure}[htbp]
		\includegraphics[width=200pt]{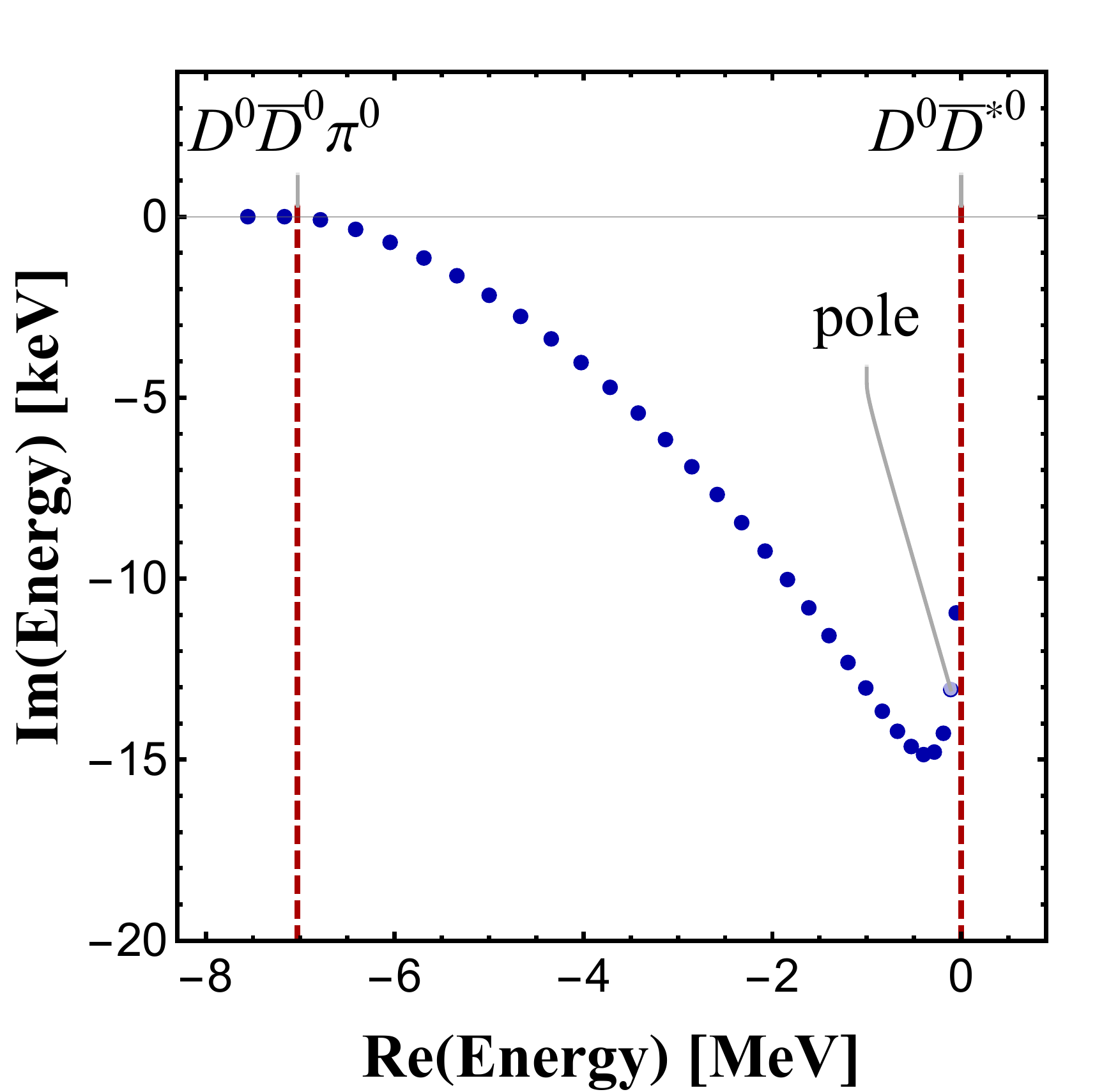}
		\caption{The dependence of the width of the $[D\bar{D}^*]$ system on the binding energy relative to the $D^{0}\bar{D}^{*0}$ threshold in the OBE potential case. The width is equal to $-2$Im(Energy), and the complex rotation angle is $\theta=15^{\circ}$. The pole in this figure corresponds to the $X(3872)$ with the $\Lambda=1155$ MeV.}\label{fig: 3872 energy width obep}
	\end{figure}

	For the $X(3872)$, the error of the binding energy given by LHCb Collaboration is significant compared with the binding energy value. Thus, we also give a list of eigenvalues with different binding energies corresponding to the pole in Fig. \ref{fig: 3872 energy width obep}. One notices that the width will vanish when the pole energy is less than the $D^{0}\bar{D}^{0}\pi^0$ threshold and reach the largest value $\approx30$ keV when the binding energy is around $-300$ keV. 
	
	\begin{widetext}
		\begin{eqnarray}
			V_{\pi}^{[D^0\bar{D}^{*0}]\to[D^0\bar{D}^{*0}]}(r)=&&\frac{g^2}{2f_\pi^2}[S(\boldsymbol{\epsilon}_3^\dagger,\boldsymbol{\epsilon}_2)Y_3(\Lambda,q_0,m_{\pi^0},r)+T(\boldsymbol{\epsilon}_3^\dagger,\boldsymbol{\epsilon}_2)H_3(\Lambda,q_0,m_{\pi^0},r)]\label{eq:OPEP r1 3872}\nonumber\\
			V_{\pi}^{[D^0\bar{D}^{*0}]\to[D^+\bar{D}^{*-}]}(r)=&&2\frac{g^2}{2f_\pi^2}[S(\boldsymbol{\epsilon}_3^\dagger,\boldsymbol{\epsilon}_2)Y_3(\Lambda,q_0,m_{\pi^+},r)+T(\boldsymbol{\epsilon}_3^\dagger,\boldsymbol{\epsilon}_2)H_3(\Lambda,q_0,m_{\pi^+},r)]\label{eq:OPEP r2 3872}\nonumber\\ 
			V_{\pi}^{[D^+\bar{D}^{*-}]\to[D^+\bar{D}^{*-}]}(r)=&&\frac{g^2}{2f_\pi^2}[S(\boldsymbol{\epsilon}_3^\dagger,\boldsymbol{\epsilon}_2)Y_3(\Lambda,q_0,m_{\pi^0},r)+T(\boldsymbol{\epsilon}_3^\dagger,\boldsymbol{\epsilon}_2)H_3(\Lambda,q_0,m_{\pi^0},r)] \label{eq:OPEP r3 3872}
		\end{eqnarray}
	\end{widetext}

	\section{Summary}\label{sec:summary}
	
	In 2021, the LHCb Collaboration observed a double-charm tetraquark $T_{cc}^{+}$ \cite{lhcbcollaborationObservationExoticNarrow2021a} with a very small binding energy relative to the $D^0D^{*+}$ threshold and a narrow width, which indicates a molecule structure. This discovery encourages us to make a careful analysis of this exotic state. We use the complex scaling method to investigate the possibility of the $T_{cc}^+$ as a $J^P=1^+$ molecule with both the OPE and OBE potential. To see the influence of the isospin breaking, we take into account the channels $D^{0}D^{*+}(^3S_1,^3D_1)$ and $D^{+}D^{*0}(^3S_1,^3D_1)$.
	
	The OPE potentials for the $DD^*$ system are unique. They have two poles located on the real transferred momentum axis. Some previous works chose the Cauchy principal value (PV) contour integral scheme, which leads to a real OPE potential. However, as shown in Fig.\ref{fig: Tcc FP}, we adopt the Feynman prescription (FP) contour integral scheme, which can provide an imaginary contribution. This imaginary part comes from the process $D^{*}\to D\pi$, which can be naturally understood in the framework of the CSM.
	
	When adopting the OBE potential, we get a pole corresponding to the $T_{cc}^{+}$, whose binding energy relative to $D^{0}D^{*+}$ is $-354$ keV, and the width is $61$ keV. The $S$-wave $D^{0}D^{*+}$ and $D^{+}D^{*0}$ components dominate the system, whose probabilities are $72.1\%$ and $27.1\%$ respectively. One can see that the isospin breaking effect plays an important role in this quasibound state. Interestingly, the OPE and OBE potential cases almost give the same results, which indicates a domination of the long-range OPE dynamics. Since our approach could give the energy and width simultaneously with only one parameter ($\Lambda$), one can predict the width once the energy is determined. Therefore, this method can provide a satisfactory description of the $T_{cc}^+$ observed in the experiment. The energy-width dependence in the $DD^*$ system is presented in Fig. \ref{fig: Tcc energy width obep}. Besides, we do not find a resonance in this framework.
	
	We also study the $X(3872)$ with the same method. Similarly, we find a quasibound state with the binding energy $-111$ keV relative to the $[D^{0}\bar{D}^{*0}]$, and the corresponding width around $26$ keV. The dominant channel is the $S-$wave $[D^{0}\bar{D}^{*0}]$ with the probability $92.7\%$. Apparently, the isospin breaking effect is significant. The numerial results are in Table \ref{tab: results}.
	
	It is helpful to make a simple analysis of the decay behavior of the $T_{cc}^+$. In this work, the $T_{cc}^{+}$ is interpreted as a molecule with the constituents $D^{0}D^{*+}$ and $D^{+}D^{*0}$. Since the channels $D^{+}D^{+}\pi^-$ and $D^{+}D^{-}\pi^+$ are kinetically forbidden, the primary decay channels are the $D^{0}D^{0}\pi^+$ and $D^{0}D^{+}\pi^{0}$. In addition, the $D^{0}D^{0}\pi^+$ is expected to be the more important channel due to the larger isospin factor. 
	
	On the other hand, the long-range OPE interactions play an important role in both $T_{cc}^{+}$ and $X(3872)$ molecules in our framework. The obtained sizes of the $T_{cc}^{+}$ and $X(3872)$ are around $5$ fm and $7$ fm respectively. Therefore, a similar long-range electromagnetic interaction may also provide a non-negligible partial width. Thus, it is very interesting to study the three-body eletromagnetic decay partial widths of the $T_{cc}^{+}$ and $X(3872)$. We may make a detailed investigation of these topics in the coming future.
	
	\acknowledgments
	{This research is supported by the National Science Foundation of China under Grants No. 11975033, No. 12070131001 and No. 12147168. The authors thank K. Chen and G. J. Wang for helpful discussions.}
	
	\bibliographystyle{apsrev4-1}
	\bibliography{tot.bib}

	
\end{document}